\documentclass{aa} 
\usepackage{amssymb}
\usepackage{natbib}
\usepackage{graphicx}
\usepackage{color}
\usepackage{txfonts}

\newcommand{\logg}{ {\rm log\,{\it g}}}
\newcommand{\teff}{ T_{{\rm eff}}}
\newcommand{\teffhydro}{ T_{{\rm eff, 3D}}}

\newcommand{\teffsol}{ T_{{\rm eff, \odot}}}
\newcommand{\teffir}{ T_{{\rm eff, IR}}}
\newcommand{\feh}{{\rm [Fe/H]}}
\newcommand{\numax}{{\nu_{\rm max}}}
\newcommand{\fbol}{{F_{\rm BOL}}}
\newcommand{\muhz}{\mu{\rm Hz}}
\newcommand{\numaxsol}{{\nu_{\rm max, \odot}}}
\newcommand{\sunmass}{{M_{\rm\odot}}}

\begin{document}
   \title{Three-dimensional interferometric, spectrometric, and planetary views of Procyon}
	\titlerunning{Three-dimensional interferometric, spectrometric, and planetary views of Procyon}
	 \authorrunning{Chiavassa et al.}

   \author{A. Chiavassa
          \inst{1}
          L. Bigot
          \inst{2}
          P. Kervella
          \inst{3}
          A. Matter
          \inst{4}
          B. Lopez
          \inst{2}
          R. Collet
          \inst{5,6}
          Z. Magic
          \inst{7}
          M. Asplund
          \inst{8}
          }

  \institute{Institut d'Astronomie et d'Astrophysique, Universit\'e Libre de Bruxelles, CP. 226, Boulevard du Triomphe, B-1050 Bruxelles, Belgium\\
              \email{achiavas@ulb.ac.be}
         \and
	Universit\'e de Nice Sophia-Antipolis, Observatoire de la C\^{o}te dÕAzur, CNRS Laboratoire Lagrange,
B.P. 4229, 06304 Nice Cedex 4, France
	\and
	LESIA, Observatoire de Paris, CNRS UMR 8109, UPMC, Universit\'e Paris Diderot, 5 place Jules Janssen, 92195 Meudon, France
	\and
	Max-Planck-Institut f\"ur Radioastronomie, Auf dem H\"ugel 69, 53121 Bonn, Germany
	\and
	Centre for Star and Planet Formation, Natural History Museum of Denmark University of Copenhagen, \O ster Voldgade 5-7, DKÐ1350 Copenhagen, Denmark
	\and
	Astronomical Observatory/Niels Bohr Institute, Juliane Maries Vej 30, DKÐ2100 Copenhagen, Denmark
	\and
	Max Planck Institute for Astrophysics, Karl-Schwarzschild-Str. 1, D-85741 Garching
	\and
	Research School of Astronomy and Astrophysics, Australian National University, Cotter Rd., Weston Creek, ACT 2611, Australia
	}
   \date{Received; accepted}

 
  \abstract
   {Procyon is one of the brightest stars in the sky and one of our nearest neighbours. It is therefore an ideal benchmark object for stellar astrophysics studies using interferometric, spectroscopic, and asteroseismic techniques.}
   {We used a new realistic three-dimensional (3D) radiative-hydrodynamical (RHD) model atmosphere of Procyon generated with the {{\sc Stagger Code}} and synthetic spectra computed with the radiative transfer code {{\sc Optim3D}} to re-analyze interferometric and spectroscopic data from the optical to the infrared. We provide synthetic interferometric observables that can be validated against observations.}
   {We compute intensity maps from a RHD simulation in two optical filters centered at 500 and 800 nm ({{\sc Mark}}~III) and one infrared filter centered at 2.2 $\mu$m ({{\sc Vinci}}). We constructed stellar disk images accounting for the center-to-limb variations and used them to derive visibility amplitudes and closure phases. We computed also the spatially and temporally averaged synthetic spectrum from the ultraviolet to the infrared. We compare these observables to Procyon data.}
   {We study the impact of the granulation pattern on center-to-limb intensity profiles and provide limb-darkening coefficients in the optical as well as in the infrared. We show how the convective related surface structures impact the visibility curves and closure phases with clear deviations from circular symmetry from the 3rd lobe on. These deviations are detectable with current interferometers using closure phases. We derive new angular diameters at different wavelengths with two independent methods based on 3D simulations. We find $\theta_{\rm{{\sc Vinci}}}=5.390 \pm 0.03$ mas and prove that this is confirmed by an independent asteroseismic estimation ($\theta_{\rm{seismic}} = 5.360 \pm 0.07$ mas. The resulting $\teff$ is 6591 K (or 6556 K, depending on the bolometric flux used), which is consistent with $\teffir=6621$ K found with the infrared flux method. We find also a value of the surface gravity $\logg=4.01 \pm 0.03$ [cm/s$^2$] that is larger by 0.05 dex from literature values. Spectrophotometric comparisons with observations provide very good agreement with the spectral energy distribution and photometric colors, allowing us to conclude that the thermal gradient of the simulation matches fairly well Procyon. \\
Finally, we show that the granulation pattern of a planet hosting Procyon-like star has a non-negligible impact on the detection of hot Jupiters in the infrared using interferometry closure phases. It is then crucial to have a comprehensive knowledge of the host star to directly detect and characterize hot Jupiters. In this respect, RHD simulations are very important to reach this aim.}
   {}


   \maketitle
%

\section{Introduction}\label{sectintro}

Procyon ($\alpha$ Canis Majoris) is one of the brightest stars in the sky and one of our nearest neighbours. It is therefore an ideal target for stellar astrophysics studies. 

For this reason, it has  a long history of observations. \citet{bessel1844} discovered that its motion was perturbed by a invisible companion. Procyon became, after Sirius, one of the first astrometric binaries ever detected. The first orbital elements were determined by \citet{auwers1862}, who showed that the period of revolution is about 40 years. The faint companion, Procyon B, was not detected visually until the end of the nineteenth century  by \citet{schaeberle1896}.  It was one of the first detected white dwarfs \citep{eggen65}.   The main component of the system is a subgiant F5 IV-V (Procyon A, HR 2943, HD 61421) which is ending its life on the Main Sequence \citep{eggenberger05, provost06}. It  has a solar metallicity \citep{griffin71,steffen85,2002ApJ...567..544A} with an effective temperature  around $\teff \approx 6500$ K \citep{code76}. One of the first radius determinations was made photometrically by \citet{gray67}, who found $R=2.24\,R_{\rm \odot}$.
The binary nature of the system is a great opportunity to determine the mass of both companions. The first attempt was made by \citet{strand51} who determined the masses of Procyon A \& B, by determining the parallax and orbital elements of the system. He found $1.76 \pm 0.1$ and $0.65 \pm 0.05 \,\sunmass$, respectively. \citet{steffen85} claimed that the mass of Procyon A was  too large to be compatible with its luminosity and suggested rather a mass around $1.4\,\sunmass$. More recent and more accurate determinations of the mass of Procyon A  converge to a star with  $M=1.497 \pm 0.037\,\sunmass$ \citep{girard00} or  $1.430 \pm 0.0034\,\sunmass$ \citep{gatewood06}. The age of the system is well constrained by the white dwarf companion, whose cooling law as function of time is well established. \citet{provencal02} found an old white dwarf with an age of $1.7 \pm 0.1$ Gy. This determination is a strong constraint for stellar evolution models.  

A way to discriminate between different masses and ages is the determination of the interferometric radius. The first attempt to measure the diameter of Procyon was done by \citet{hanbury67,1974MNRAS.167..475H} who found an angular diameter of $\theta = 5.50 \pm 0.17$ mas. This value was confirmed later by \citet{1991AJ....101.2207M} but with a much better precision ($\sim 1$\%). More recently, \citet{2004A&A...413..251K} redetermined the angular diameter using the {\sc Vinci} instrument at VLTI. They found a diameter that is even smaller $\theta = 5.448 \pm 0.053$ mas. \citet{2005ApJ...633..424A} re-analyzed these data using hydrodynamical model atmospheres and found $\theta=5.404 \pm 0.031$ mas (see their Table 7).
It is interesting to compare these angular diameters  with the independent infrared flux method, a recent determination  was performed by \citet{2010A&A...512A..54C} who derived a value of $\theta_{\rm IR} = 5.326 \pm 0.068$ mas, which is smaller than the \citeauthor{2003A&A...408..681K}'s result, but they agree to within 1$\sigma$. 

Another great particularity of Procyon is the presence of oscillations that are due to trapped acoustic modes. The first claims of a detection of  an excess power were made by \citet{gelly86,gelly88} and \citet{brown91} who found a mean large spacing between consecutive acoustic modes of about $39\,\muhz$ and $55\,\muhz$, respectively. \citet{Martic99} made a clear detection using the ELODIE fiber \'echelle spectrograph with a strong excess power around $\numax \approx 1000\,\muhz$ and confirmed the results of \citet{brown91}. They determined the frequency spacing to be $55\,\muhz$.   Later,  \citet{eggenberger04} and \citet{martic04} made the first identifications of individual frequencies of spherical harmonic degrees $\ell=0,1,2$ with mean large spacings of $55.5\pm 0.5$ \citep{eggenberger04} and $53.5 \pm 0.5 \,\muhz$ \citep{martic04}. More recent observations have been made from the ground during single or multisite  campaigns \citep{mosser08, arentoft08,bedding10} or from space on board of the MOST satellite \citep{matthews04,guenther08}, which considerably improved the precision in frequencies within $\sim 1\,\muhz$.

The stellar evolution model of Procyon was made by \citet{hartmann75}, who used the astrometric mass, photometry, and the angular diameter  of \citet{1974MNRAS.167..475H} to  constrain the model. More realistic modeling, especially with better equation-of-states, came with \citet{guenther93}, who showed the importance of diffusion of elements. The values of individual p-modes frequencies considerably  constrained the determinations of the fundamental parameters \citep{barban99,dimauro01,eggenberger05,provost06,bonanno07,guenther08}. 
\citet{eggenberger05} and \citet{provost06} made a realistic stellar evolution modeling to constrain simultaneously the location in the HR diagram and the large frequency separations of \citet{eggenberger04} and \citet{martic04}, respectively. They disagreed on the derived mass. \citeauthor{eggenberger05} found a mass that agrees well with the astrometric value derived by \citet{girard00}, whereas \citeauthor{provost06} found a mass that corresponds to the more recent value of \citet{gatewood06}. Part of the difference in the two stellar evolution models comes from the slightly different large separation (which increases with the mass), and partly due to the difference in the stellar evolution codes themselves. We emphasize that the best stellar model of \citeauthor{provost06} that fits asteroseismic and spectro-photometric data has an age ($\geq 2$ Gy) that is consistent with the age of the white dwarf.  Indeed, the astrometric mass of Procyon B implies that the mass of its  progenitor was about $3 \sunmass$, which has lived about  $\sim 500$ My on the Main Sequence. Therefore, we can safely conclude that the age of the system  must be at least 2 Gy.  The larger mass of \citet{eggenberger05} corresponds to a younger ($\leq1.7$ Gy) system, incompatible with the result of \citet{provencal02}. 
 
The atmospheric parameters ($\teff, \logg, \feh$) and the interferometric radius, that are used to define the stellar evolution model and the analysis of frequencies, strongly depend on the realism of the atmosphere and the exactness  of the temperature gradient in the surface layers. 
In the case of F-stars, these gradients are strongly modified by the convective transport which is more vigorous than in the Sun. This is clearly seen in line bisectors \citep{gray81,dravins87}, which are two or three times larger than in the Sun. 
The higher convective velocities are due to the higher stellar luminosity and smaller densities.  This strong effect of convection must be taken into account for stellar physics diagnostics. Realistic 3D time-dependent hydrodynamical simulations of the surface layer of Procyon were performed by several authors \citep{atroshchenko89,nordlund90,2002ApJ...567..544A}, who showed  that the 3D effects in such F-star can bring significant differences for line profile formation and abundance analysis \citep{2002ApJ...567..544A}. \citet{nelson80} and \citet{nordlund90} showed that the surface defined at $T=\teff$ is not flat but rather ``corrugated'' due to the large fluctuations and the high contrast of granulation.  These hydrodynamical simulations can reproduce with success the line shifts, asymmetries, and in particular the observed bisectors of various lines  \citep[][ FeI and FeII]{2002ApJ...567..544A}.
\citeauthor{2002ApJ...567..544A} also showed that the 3D limb darkening law significantly differ 1D law, up to $\sim1.6\%$, leading to a correction of $\Delta \teff \approx 50$ K, which is non-negligible for precise stellar evolution modeling.  \citet{2005ApJ...633..424A}  made 3D models using the CO$^5$BOLD code \citep{2002AN....323..213F,2011arXiv1110.6844F} to calculate limb darkened intensity profiles to analyze visibility curves obtained by the {\sc Vinci} instrument \citep{2003A&A...408..681K} and the {\sc Mark}~III. Their 3D analysis led to a smaller radius by $0.04$ mas than that of \citet{2003A&A...408..681K}, who used 1D limb darkened law. \citeauthor{2005ApJ...633..424A} also showed that 3D models better reproduce the spectral energy distribution in the UV, whereas 1D model are unable.

Regarding the importance of a realistic hydrodynamical modeling of the atmosphere of Procyon, we re-analyze the interferometric and spectroscopic data at the different wavelengths of \citet{2005ApJ...633..424A} using up-to-date line and continuum opacities \citep{2008A&A...486..951G} to derive a new radius. We propose a solution that agrees well with the astrometric, asteroseismologic, infrared flux method, and interferometric data. We also explore the impact of convection-related surface structures on the closure phases and assess how the direct search of planets by interferometry may be affected by the host star surface structures.

\section{Three-dimensional radiative-hydrodynamical approach}

\subsection{Procyon simulation}

The convective surface of Procyon is modeled  using the {\sc Stagger Code}  \citep[][Nordlund \& Galsgaard\footnote{1995,
http://www.astro.ku.dk/$\sim$kg/Papers/MHDcode.ps.gz}]{2009LRSP....6....2N}. 
In a local box located around the optical surface $\tau \approx 1$,  the code solves the full set of hydrodynamical equations for the conservation of mass, momentum, and energy
coupled to an accurate treatment of the radiative transfer.  The equations are solved on a staggered mesh where the thermodynamical scalar variables (density, internal energy, and temperature) are cell centered, while the fluxes are defined on the cell faces.  This scheme has several numerical advantages to simulate surface
convection. It is robust against shocks and ensures conservation of the thermodynamic variables.  The domain of simulation contains the entropy minimum located at the surface and is extended deep enough to have a flat entropy profile at the bottom. The code uses periodic boundary conditions horizontally and open boundaries
vertically. At the bottom of the simulation, the inflows have constant entropy and
pressure. The outflows are not constrained and are free to pass through the
boundary. The code is based on a sixth order explicit finite difference scheme and fifth order for interpolation.
Numerical viscosity of the Rytchmeyer \& Morton type is used to stabilize the  code. The corresponding adjustable parameters are chosen to minimize
the viscosity and are not adjusted to fit the observables.  We used a realistic equation-of-state that accounts for ionization, recombination,
and dissociation \citep{MHD} and continuous (Trampedach et al. private communication) and line opacities \citep{2008A&A...486..951G}.
An accurate treatment of the transfer is needed to get a correct temperature
gradient of the transition region between optically thin and thick layers. 
The transfer equation is solved using a Feautrier-like scheme along several inclined rays (one vertical,
eight inclined) through each grid point. The wavelength dependence of the radiative
transfer is taken into account using opacity bins \cite{1982A&A...107....1N}. The numerical resolution  is $240^3$. The geometrical sizes are 22 Mm $\times$ 22 Mm horizontally and 17
Mm vertically. The horizontal dimensions of the box are defined to include at each time step a sufficient number of
granules and the vertical one is chosen to ensure that the entropy profile is flat at the bottom. The equations of magnetohydrodynamics are not computed for this model. The stellar parameters corresponding to our RHD model (Table~\ref{simus}) are $\teff = 6512 \pm 25$ K, $\logg
= 4.0$ [cm/s$^2$], and a solar chemical composition \citep{asplund09}.  The uncertainty in $\teff$ represents the fluctuations with time around the
mean value. These parameters roughly correspond to those of Procyon. The exact values of the parameters do not influence the limb darkening
\citep{2005ApJ...633..424A, 2011A&A...534L...3B}.

\subsection{Spherical tiling models, intensity maps, and spectra}\label{tilingsect}

 \begin{figure}
   \centering
   \includegraphics[width=0.91\hsize]{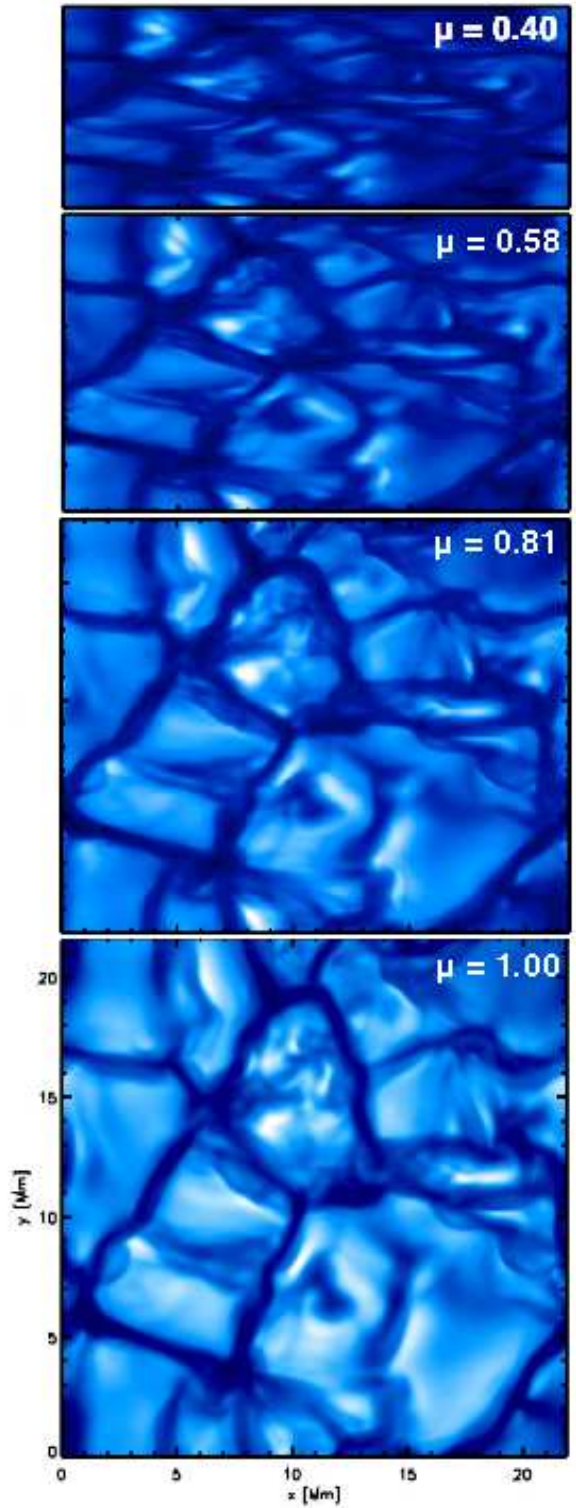}
    \caption{Intensity maps in the {\sc Mark} 500 nm filter for the RHD simulation for different inclination angles $\mu$. The intensity ranges from $1{\times}10^3$ to $1.1\times10^7$\,erg\,cm$^{-2}$\,s$^{-1}$\,{\AA}$^{-1}$.}
              \label{limbfig}%
    \end{figure}
   
The computational domain of each simulation represents only a small portion of the stellar surface. To obtain an image of the whole stellar disk, we employ the same tiling method explained in \cite{2010A&A...524A..93C}. For this purpose, we used the 3D pure-LTE radiative transfer code {\sc Optim3D} \citep{2009A&A...506.1351C} to compute intensity maps from
the snapshots of the RHD simulation of Table~\ref{simus} for different inclinations with respect to the vertical, $\mu{\equiv}\cos(\theta)$=[1.000, 0.989, 0.978, 0.946, 0.913, 0.861, 0.809, 0.739, 0.669, 0.584, 0.500, 0.404, 0.309, 0.206, 0.104] (Fig.~\ref{limbfig}) and for a representative series of simulation's snapshots: we chose $\sim$ 25 snapshots taken at regular intervals and covering $\sim$1 h of stellar time, which corresponds to $\sim$5 p-modes. {\sc Optim3D} takes into account the
Doppler shifts due to convective motions and the radiative
transfer equation is solved monochromatically using pre-tabulated extinction coefficients as functions of temperature, density, and
wavelength (with a resolving power of $\lambda\slash\delta\lambda=500\,000$). The lookup tables were computed for the same chemical composition as the RHD simulation \citep[i.e.][]{asplund09} and 
using the same extensive atomic and molecular opacity data as the latest generation of
{\sc MARCS} models \citep{2008A&A...486..951G}.\\
Then, we used the synthetic images to map onto spherical surfaces accounting for distortions especially at high latitudes and longitudes cropping the square-shaped intensity maps when defining the spherical tiles. Moreover, we selected intensity maps computed from random snapshots in the simulation time-series: this process avoided the assumption of periodic boundary conditions resulting in a tiled spherical surface globally displaying an artifactual periodic granulation pattern.

Based on the stellar radius estimates and on the sizes of the simulationsÕ domains (Table \ref{simus}), we required 215 tiles to cover half a circumference from side to side on the sphere (number of tiles $= \pi \cdot R_\star / 22.0 $, where 22.0 is the horizontal dimension of the numerical box in Mm and $R_\star$ the radius of the star). To produce the final stellar disk images, we performed an orthographic projection of the tiled spheres on a plane perpendicular to the line- of-sight ($\mu=1.0$). The orthographic projection returned images of the globe in which distortions are greatest toward the rim of the hemisphere where distances are compressed \citep{2010A&A...524A..93C}.

\begin{table*}
\centering
\begin{minipage}[t]{\textwidth}
\caption{3D simulation of Procyon used in this work.}             
\label{simus}      
\centering                          
\renewcommand{\footnoterule}{} 
\begin{tabular}{c c c c c c}        
\hline\hline                 
$<T_{\rm{eff}}>$\footnote{Horizontally and temporal average and standard deviation of the emergent effective temperatures} & [Fe/H]  & $\log g$ & $x,y,z$-dimensions & $x,y,z$-resolution   & $\rm{R}_{\star}$ \\
$[\rm{K}]$ & & [cm/s$^2$]  & [Mm]  & [grid points]   & [$\rm{R}_\odot$]\\
\hline
6512$ \pm 25$\footnote{\cite{2011JPhCS.328a2003C}} & 0.0\footnote{Chemical composition by \cite{asplund09}} & 4.0 &  22.0$\times$22.0$\times$17.0  & 240$\times$240$\times$240 & 2.055\footnote{Angular diameter of 5.443 mas \citep{2004A&A...413..251K} converted into linear radius with Eq.~(\ref{eq_seismic})} \\
\hline\hline                          
\end{tabular}
\end{minipage}
\end{table*}

\begin{figure}
   \centering
   \begin{tabular}{ccc}
        \includegraphics[width=0.98\hsize]{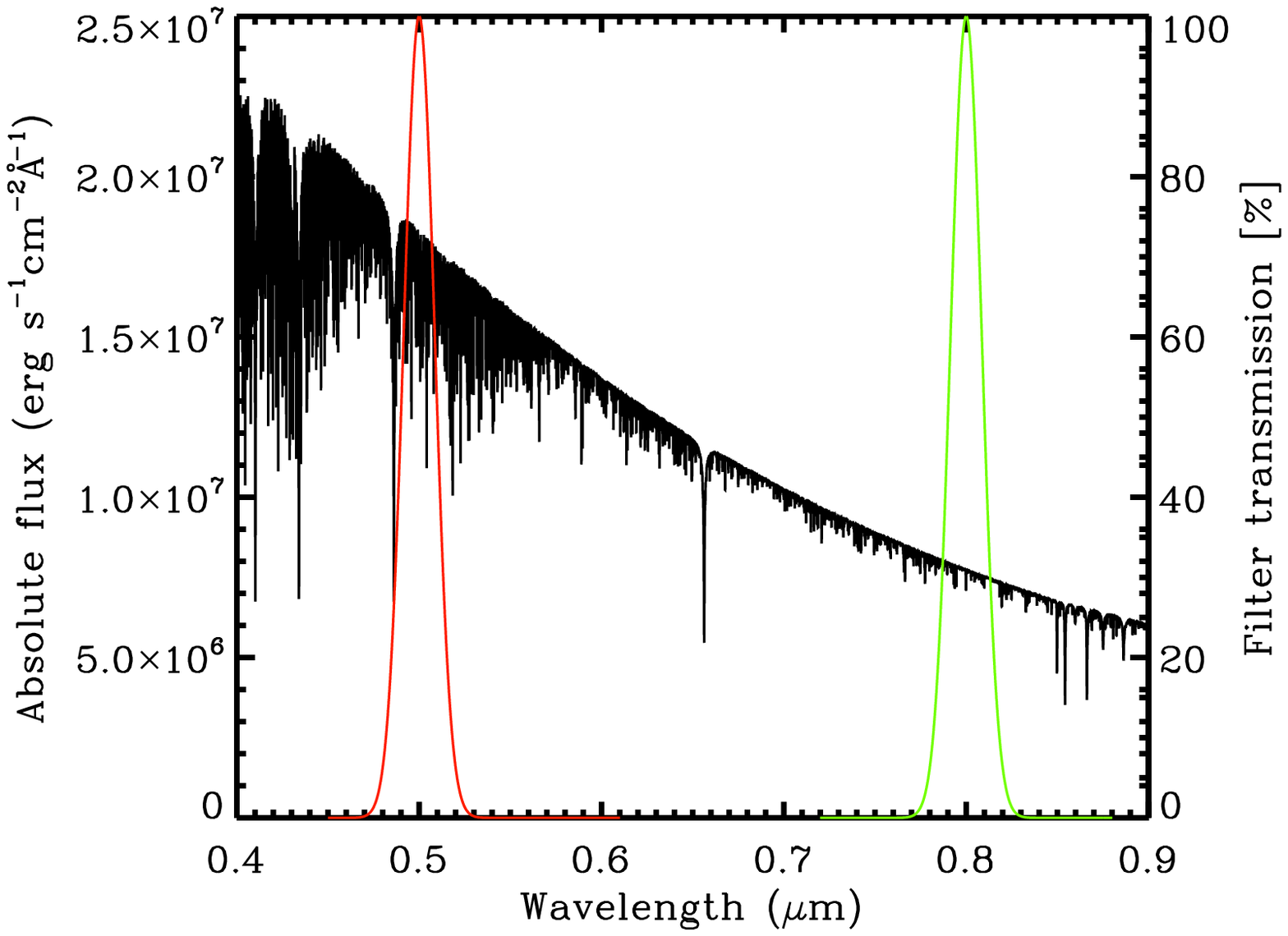}\\
           \includegraphics[width=0.98\hsize]{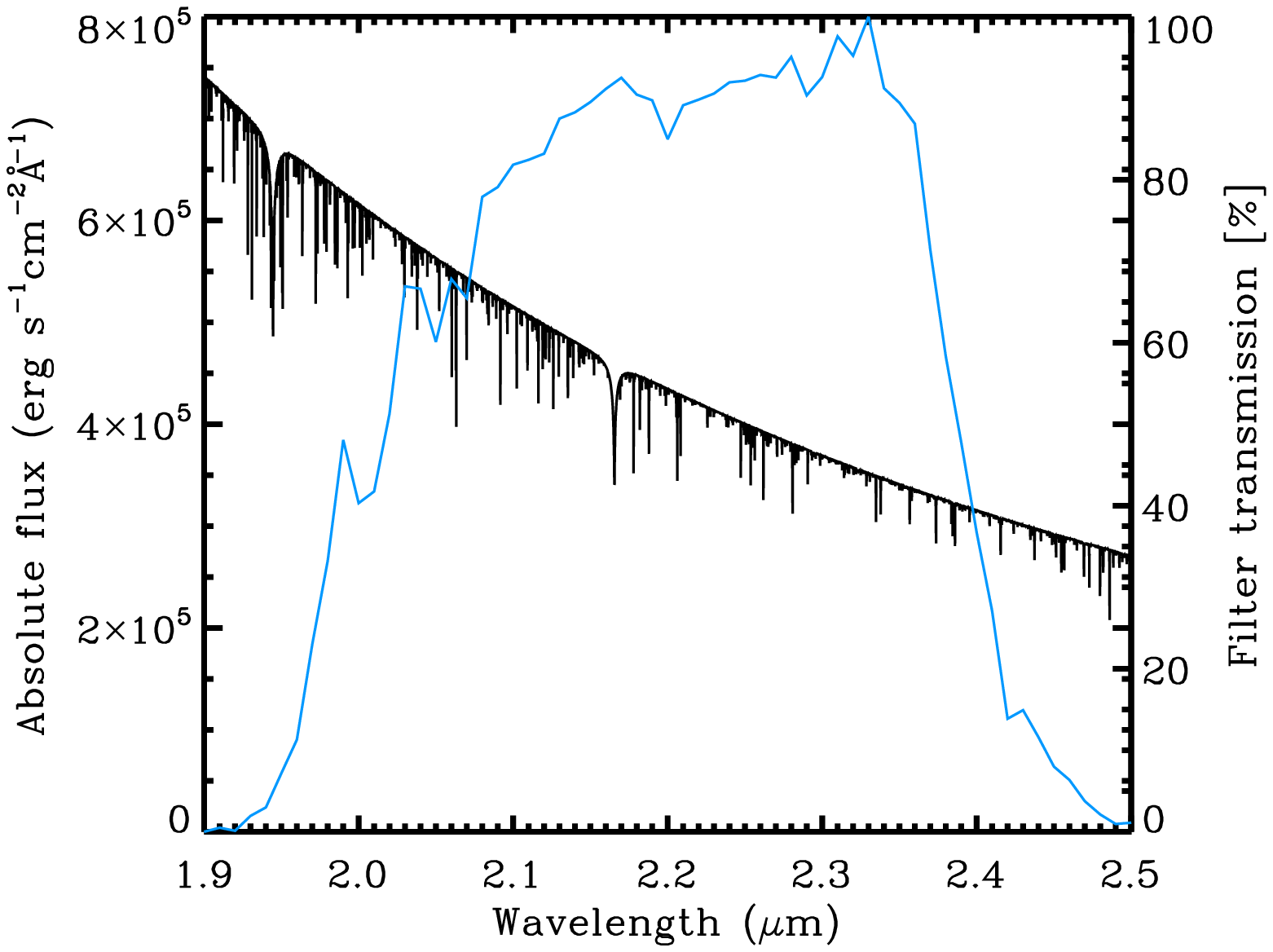}
        \end{tabular}
      \caption{A spatially and temporally averaged synthetic spectrum in the optical wavelength region with the {\sc Mark} 500 nm filter (red: top left), {\sc Mark} 800 nm (green: top right), and {\sc Vinci} infrared filter (blue: bottom).}
        \label{filters}
   \end{figure}

\begin{figure*}
   \centering
   \begin{tabular}{ccc}
   \includegraphics[width=0.3\hsize]{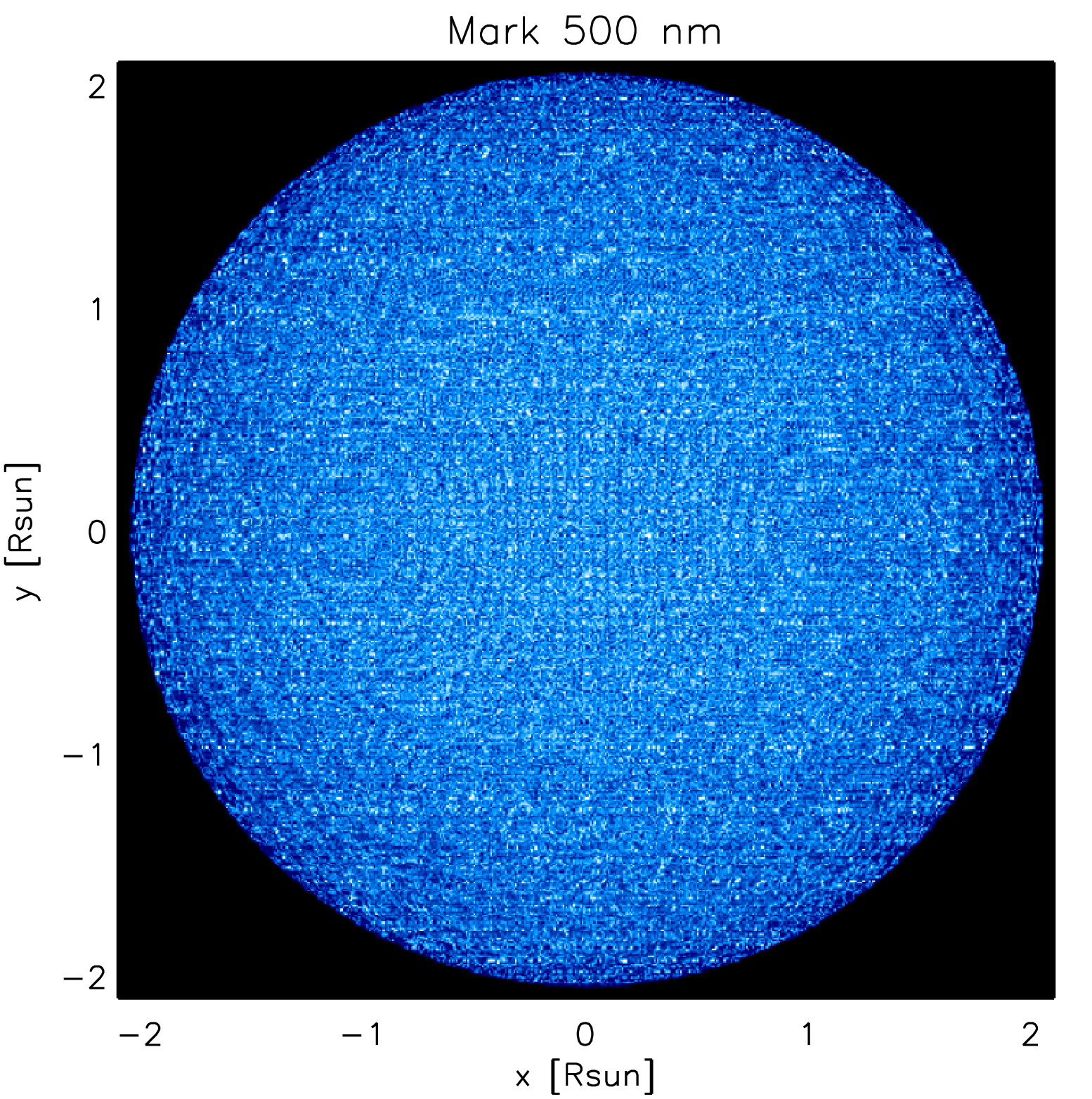}
       \includegraphics[width=0.3\hsize]{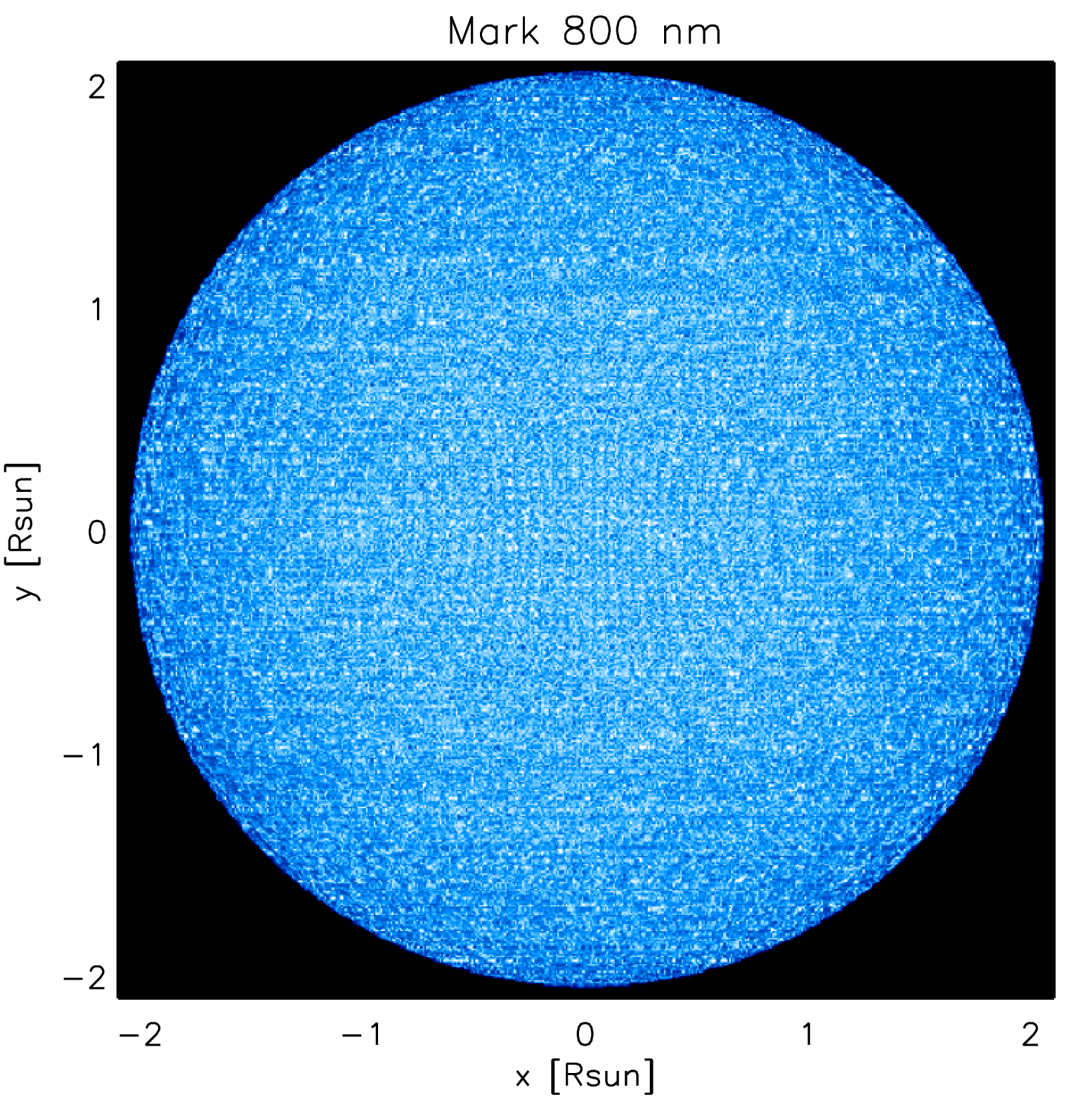}
           \includegraphics[width=0.3\hsize]{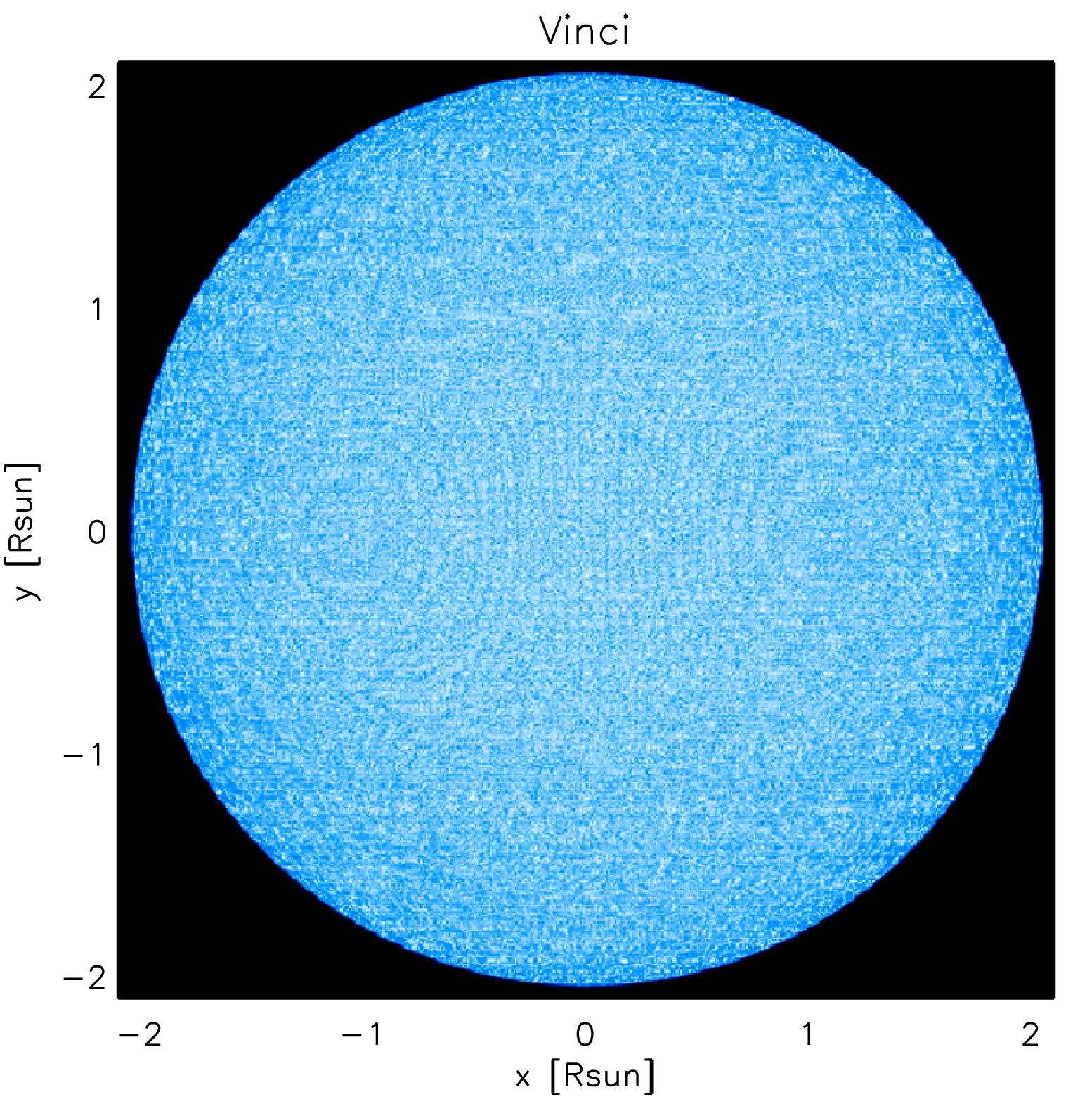}\\
              \includegraphics[width=0.3\hsize]{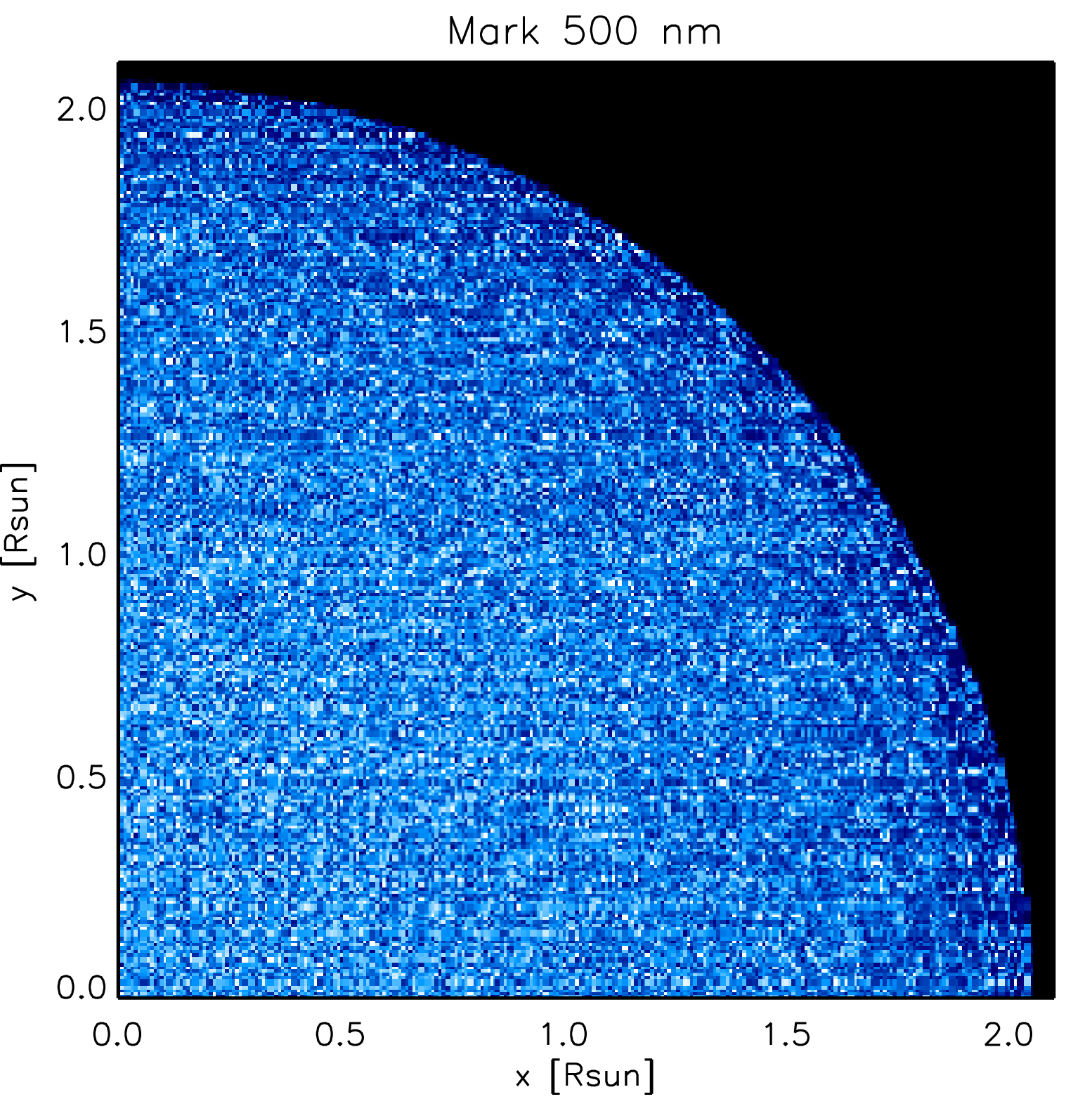}
       \includegraphics[width=0.3\hsize]{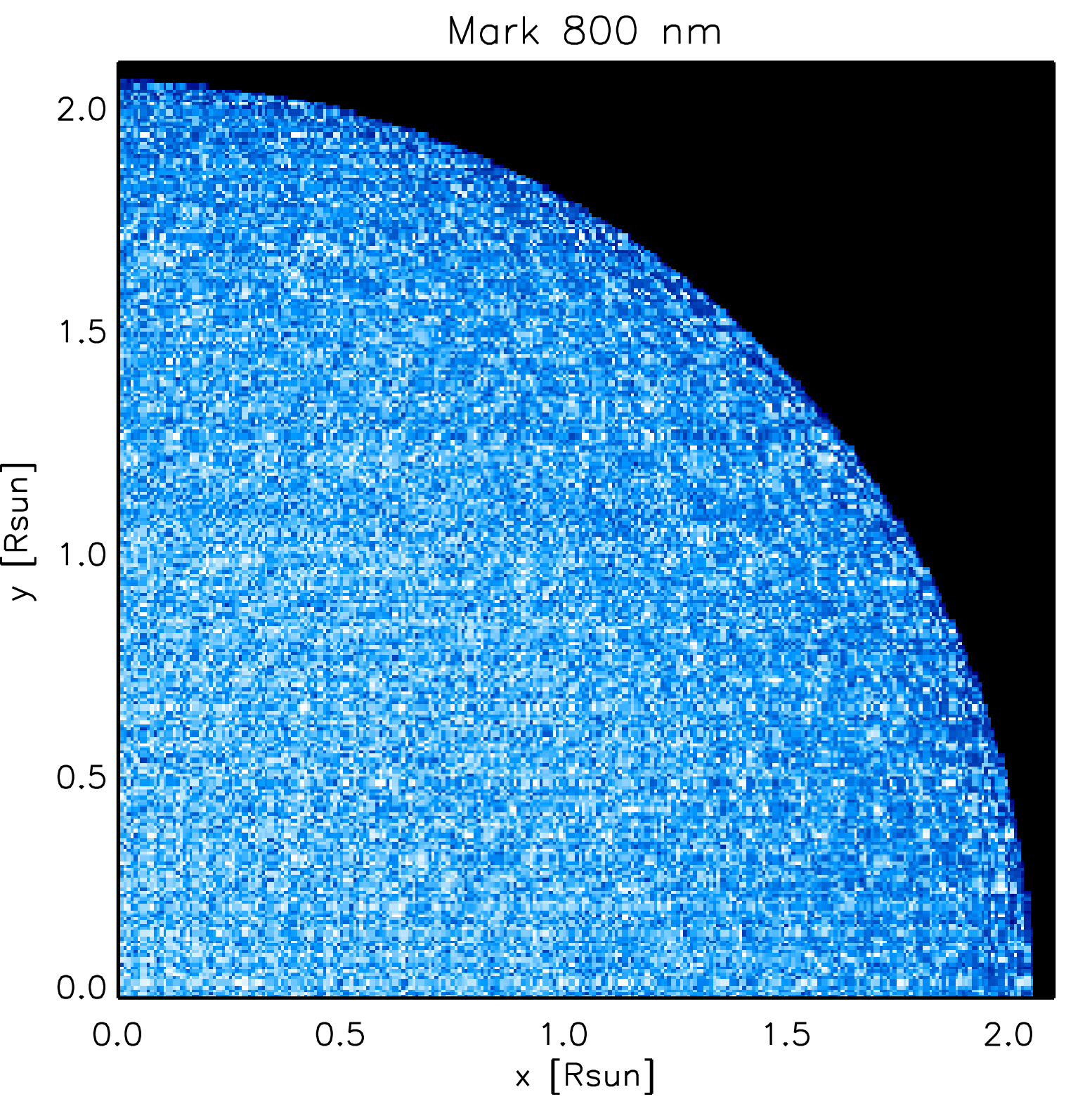}
           \includegraphics[width=0.3\hsize]{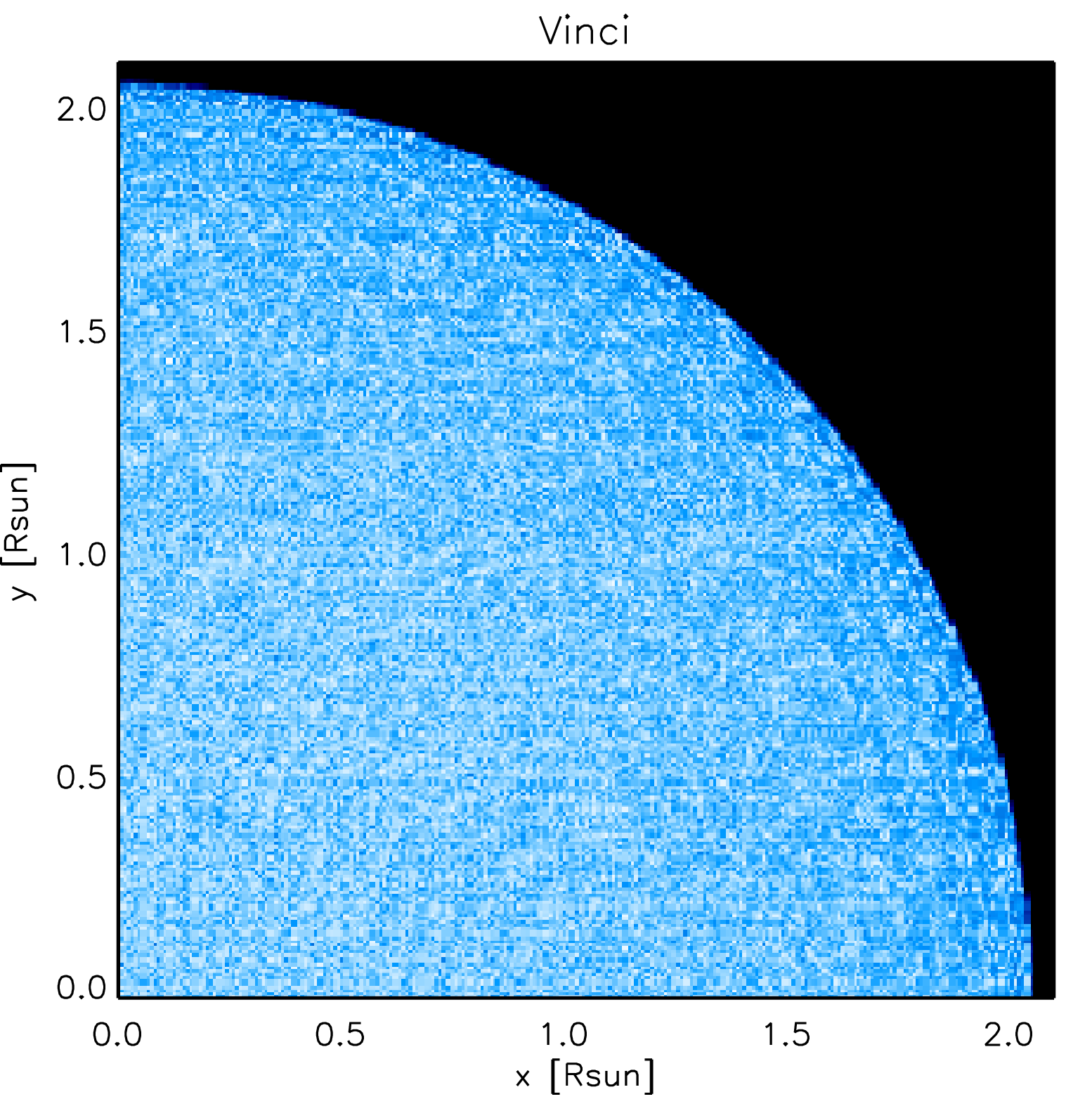}
        \end{tabular}
      \caption{\emph{Top row:} synthetic stellar disk images of the RHD simulation. The intensity range is [0.3 - 1.0$\times10^6$], [0.3 - 5.2$\times10^5$], and [0.3 - 0.9$\times10^5$]\,erg\,cm$^{-2}$\,s$^{-1}$\,{\AA}$^{-1}$ for {\sc Mark} 500 nm, {\sc Mark} 800 nm, and {\sc Vinci} filters, respectively. \emph{Bottom row:} enlargements of the  images above.}
        \label{images}
   \end{figure*}

In this work, we computed a synthetic stellar disk image for interferometric spectral bands used in \cite{2005ApJ...633..424A}: (i) the {\sc Mark}~III \citep{1988A&A...193..357S} centered at 500 and 800 nm, and (ii) {\sc Vinci} \citep{2003SPIE.4838..858K} centered at 2.2 $\mu$m (Fig.~\ref{filters}). The {\sc Mark}~III sensitivity curves are assumed to be Gaussian with central wavelengths $\lambda_0=$ 500 and 800 nm, each of them with a FWHM of 20 nm \citep{1991AJ....101.2207M}. We produced a number of synthetic stellar disk images corresponding to different wavelengths in the filters with a spectral resolving power of 20\,000. Figure~\ref{images} shows the resulting synthetic stellar disk images averaged over each passband.\\
With {\sc Optim3D}, we computed also the spectra, which are normalized to the filter transmission as: $\frac{\int I_{\lambda} T\left(\lambda\right)d\lambda}{\int T\left(\lambda\right)d\lambda}$, where 
$I_\lambda$ is the intensity and $T\left(\lambda\right)$ is the transmission curve of the filter 
at a certain wavelength. The spectra in Fig.~\ref{filters} have been
computed along rays of four $\mu$-angles [0.88, 0.65, 0.55, 0.34] and four $\phi$-angles [$0^{\circ}$,
$90^{\circ}$, $180^{\circ}$, $270^{\circ}$], after which we
performed a disk integration and a temporal average over all selected
snapshots. 

\subsection{Three-dimensional limb-darkening}

Figure~\ref{images} shows irregular stellar surfaces with numerous convective related surface structures. There are pronounced center-to-limb variations in the {\sc Mark} 500 nm and {\sc Mark} 800 nm filters while these are less noticeable in the {\sc Vinci} filter. This is manly due to the different Planck functions in the optical range and in the infrared region.

\begin{figure*}
   \centering
   \begin{tabular}{ccc}
   \includegraphics[width=0.33\hsize]{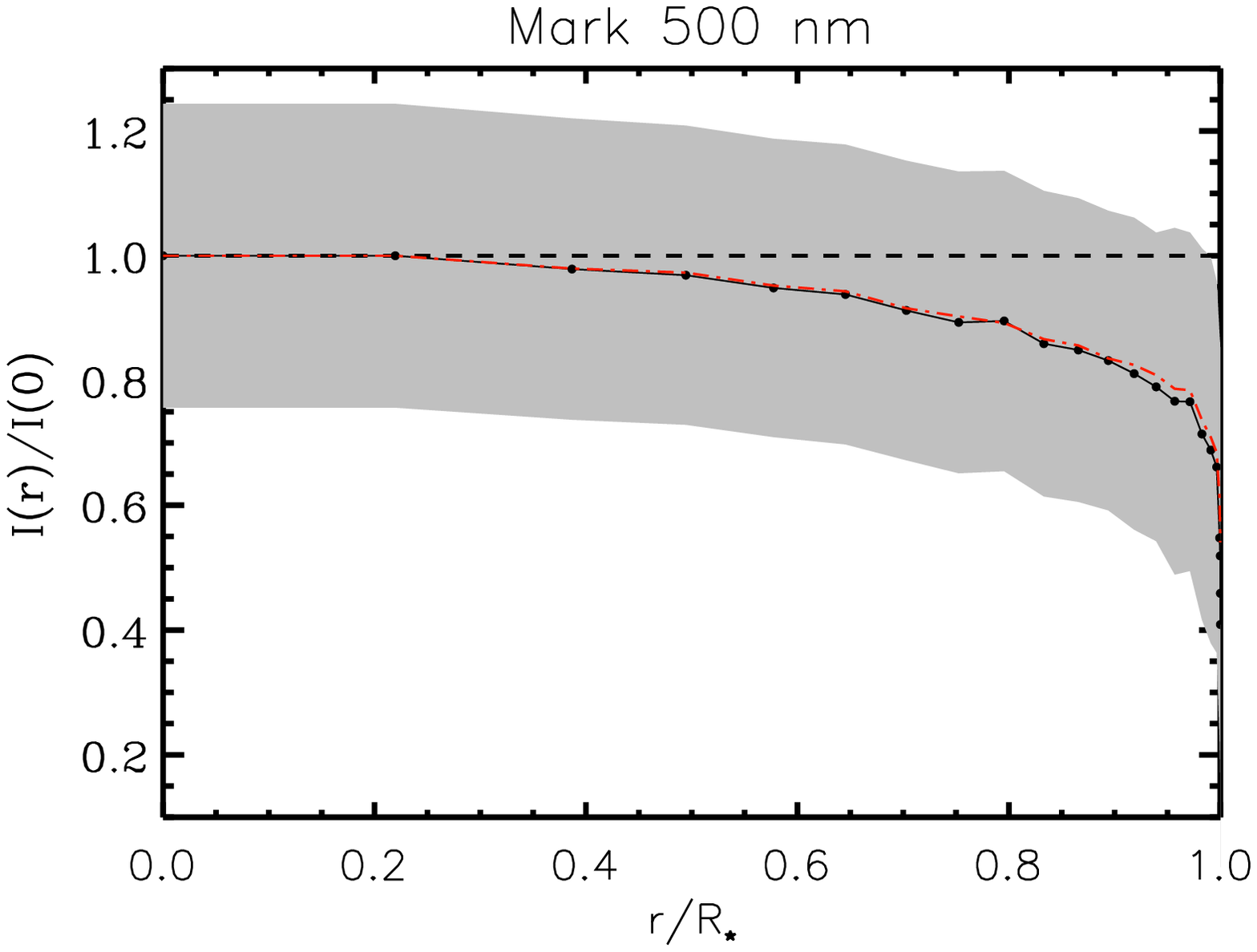}
          \includegraphics[width=0.33\hsize]{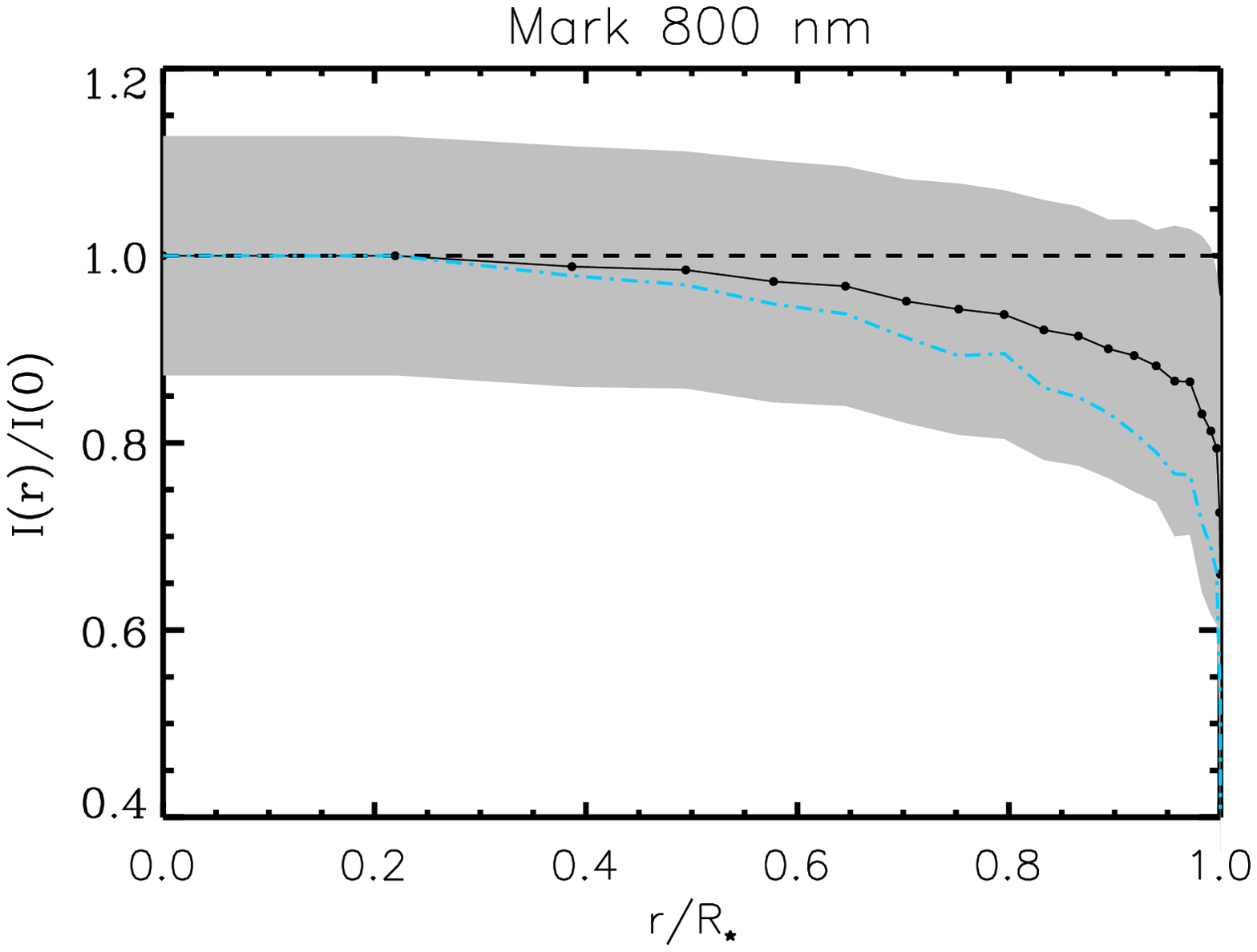}
           \includegraphics[width=0.33\hsize]{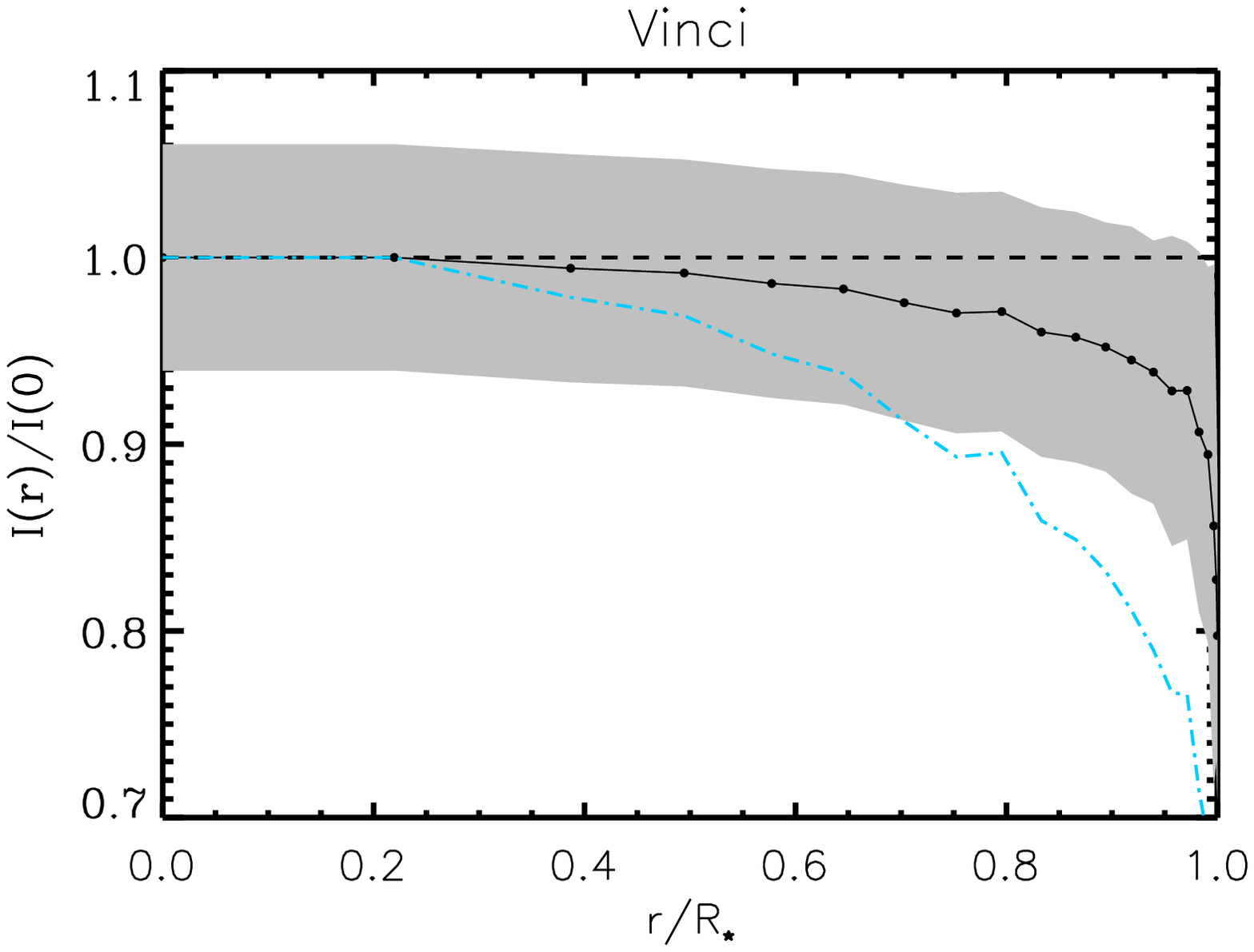}
        \end{tabular}
      \caption{Radially averaged intensity profiles (black line) derived from the synthetic stellar disk images of Fig.~\ref{images}. The gray areas denote the one sigma spatial fluctuations with respect to the averaged intensity profile. The dashed horizontal line is an uniform disk intensity profile. The intensity is normalized to the mean intensity at disk center and the radius is normalized to the radius given in Table~\ref{simus}. The blue dotted-dashed line in \emph{central} and \emph{right} panels is the average intensity profile of {\sc Mark} 500 nm. The red dotted-dashed line in \emph{left} panel is the continuum only averaged intensity profile in {\sc Mark} 500 nm.}
        \label{profiles}
   \end{figure*}

We derived azimuthally averaged (i.e., averaged over different $\phi$ angles) intensity profiles for every synthetic stellar disk image from the simulation (Fig.~\ref{profiles}). Using the method described in \cite{2009A&A...506.1351C,2010A&A...524A..93C}, the profiles were constructed using rings regularly spaced in $\mu=\cos(\theta)$ for $\mu\le1$ (i.e. $r/\rm{R}_{\star}\le1$), with $\theta$ the angle between the line of sight
and the vertical direction. The standard deviation of 
the average intensity, $\sigma_{I\left(\mu\right)}$, was computed within each ring, the $\mu$ parameter is connected to the impact parameter $r/\rm{R}_{\star}$ through the relationship $r/\rm{R}_{\star}=\sqrt{1-\mu^2}$, where $\rm{R}_{\star}$
is the stellar radius reported in Table~\ref{simus}. The total number of rings is 20, we ensured that this number is enough to have a good characterization of the intensity profiles. \\
Figure~\ref{profiles} displays a steeper center-to-limb variation for the optical region, as already visible in the disk images, with fluctuations of $\sim20\%$ in the {\sc Mark}~III 500 nm filter down to $\sim10\%$ and $\sim5\%$ in the {\sc Mark}~III 800 nm and {\sc Vinci} filters, respectively. We tested the impact of spectral lines in the {\sc Mark} 500 nm filter, for which the effects of lines are stronger, computing a synthetic disk image considering only the continuum opacities. Its averaged intensity profiles (red line in left panel of Fig.~\ref{profiles}) is very similar to the one computed with spectral lines (black line) with differences lower than $1\%$ for $r/\rm{R}_{\star}\le0.9$ and between $1-5\%$ at the limb ($0.9<r/\rm{R}_{\star}\le1.0$). It is also interesting to notice that the continuum profile tends to be closer to the uniform disk profile (dashed line in the figure), as well as the {\sc Mark} 800 nm and {\sc Vinci} filters with respect to the intensity profile of the {\sc Mark} 500 nm.

\begin{figure}[h!]
   \centering
   \begin{tabular}{ccc}
   \includegraphics[width=0.98\hsize]{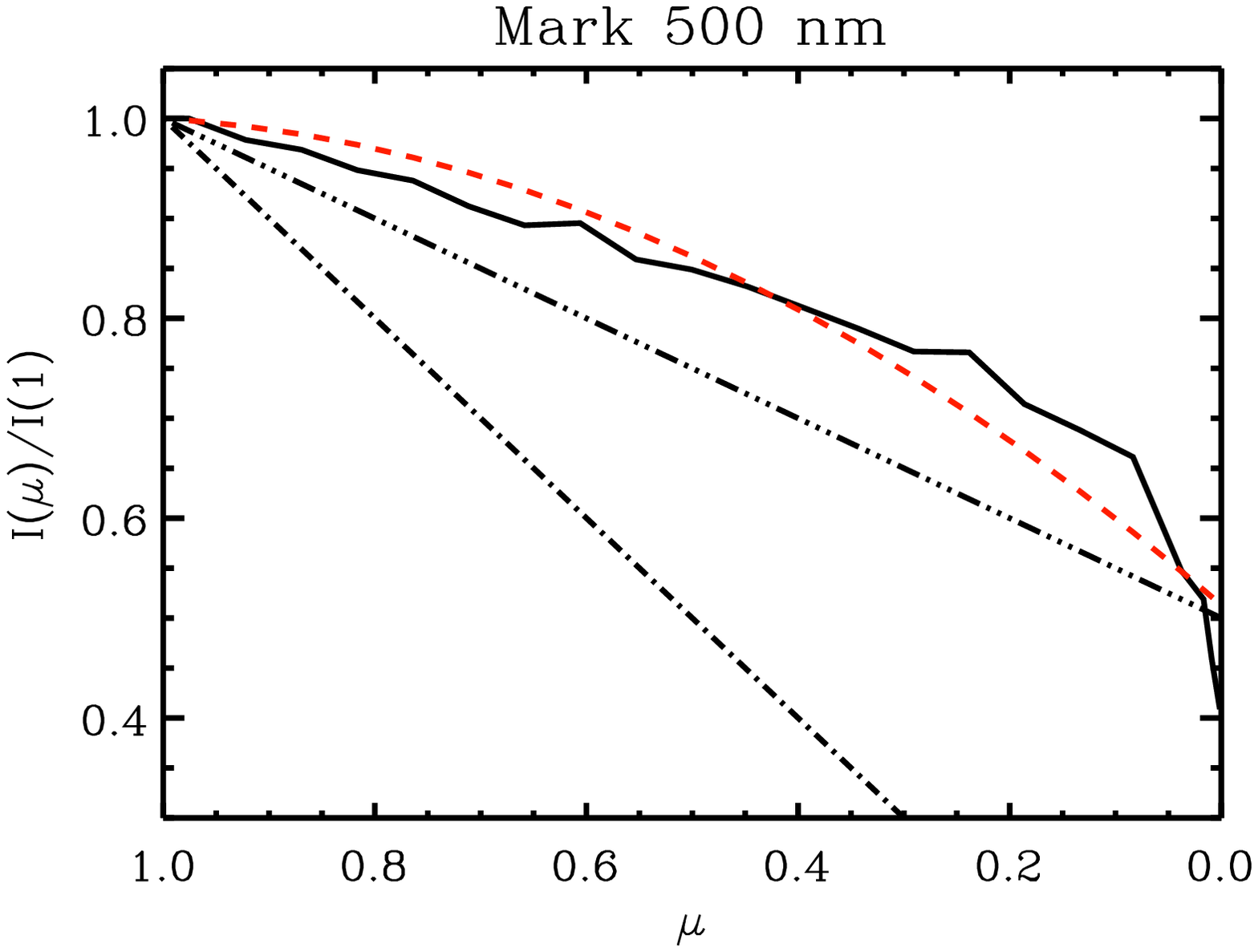}\\
       \includegraphics[width=0.98\hsize]{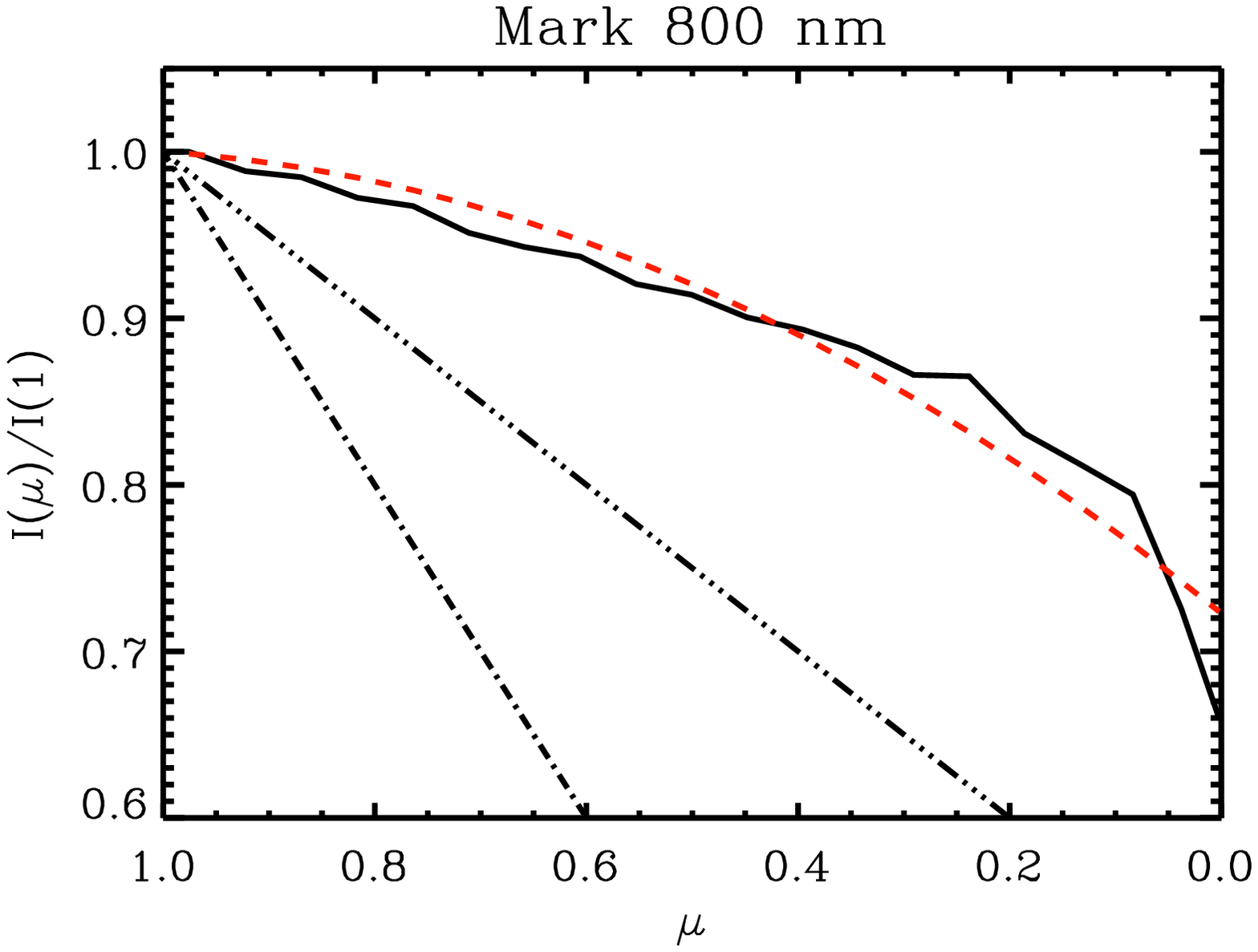}\\
           \includegraphics[width=0.98\hsize]{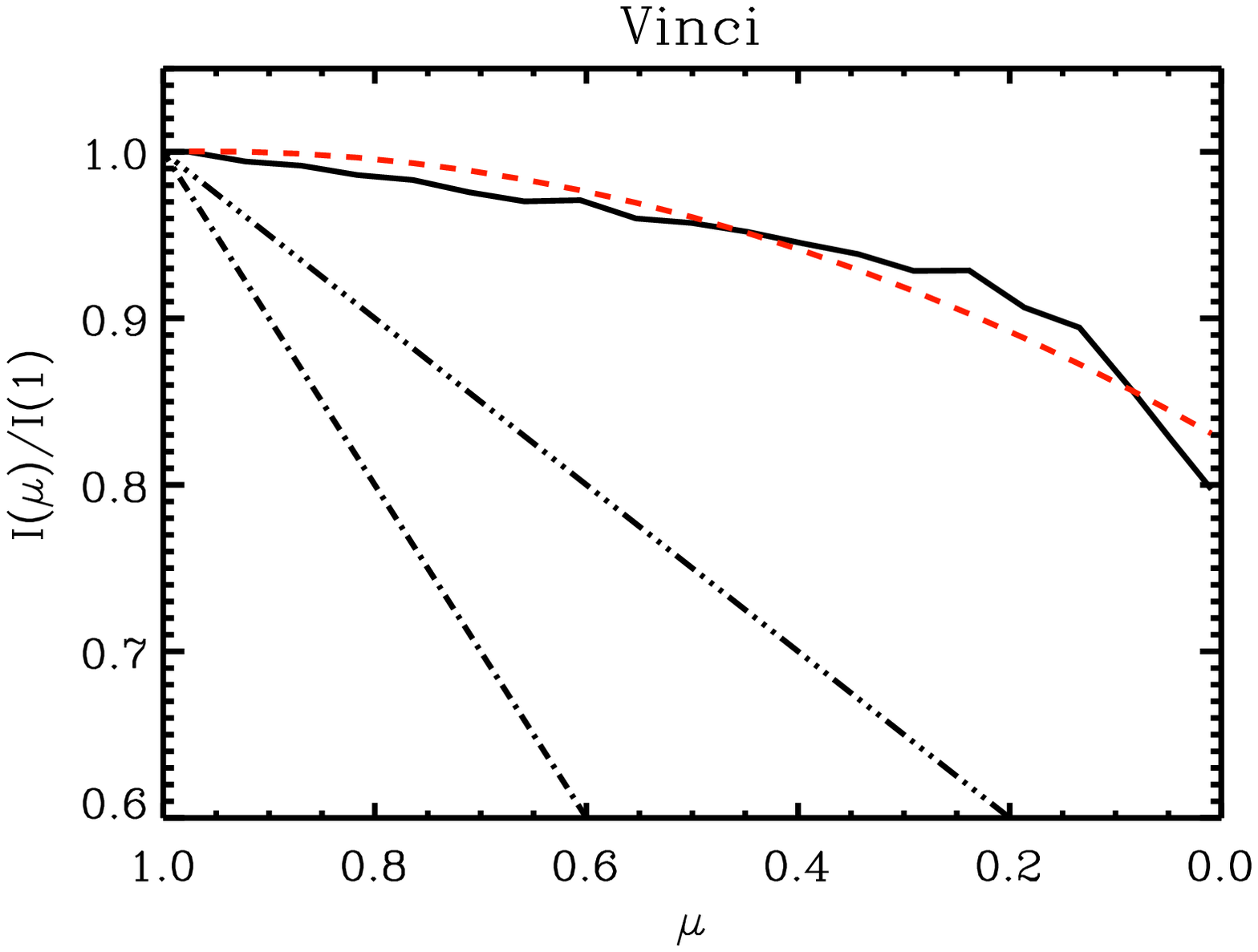}
        \end{tabular}
      \caption{Center-to-limb fits (red dashed line) obtained using Eq.~(\ref{claret_law}) with $N=2$ for the 3D azimuthally average intensity profile (solid line). A full limb-darkening (dash-dotted line), a partial limb-darkening (triple dot-dashed line) are also shown. $\mu$ = $\cos\theta$, with $\theta$ the angle between the line of sight and the vertical direction.} 
        \label{profiles_fits}
   \end{figure}

We used the following limb-darkening law \citep{2009A&A...506.1351C} to fit the averaged profiles: 

\begin{equation}\label{claret_law}
\frac{I_{\lambda}(\mu)}{I_{\lambda}(1)}=\sum_{k=0}^N a_k\left(1-\mu\right)^k.
\end{equation}

In this equation, $I_{\lambda}(\mu)$ is the intensity, $a_k$ are the limb-darkening coefficients, and $N+1$ their number.  We performed a Levenberg-Marquardt
least-square minimization to fit all the radially averaged profiles of Fig.~\ref{profiles} using this law and weighting the fit by $1/\sigma_{I\left(\mu\right)}$, due to 3D fluctuations (gray areas of Fig.~\ref{profiles}) We varied the order $N$ and found that $N=2$ provides the optimal solution with very minor improvements  to the $\chi^2$ minimization using $N=3$. This was already found by \cite{2010A&A...524A..93C} for K giants. Figure~\ref{profiles_fits} shows the fits in the different filters used in this work, while Table~\ref{LDcoeff} reports the limb-darkening coefficients from the fits. The figure illustrates that the average
intensity profiles from a RHD simulation is largely different from the full limb-darkening
($I(\mu)/I(1)=\mu$) and partial limb-darkening ($I(\mu)/I(1)=0.5\cdot\mu$) profiles as well as the power law profile that does not even provide an appreciable fit to the radially average profile. We therefore discourage the use of these simple laws.

\begin{table}
\begin{minipage}[t]{\columnwidth}
\caption{Limb-darkening coefficients from the law described in text for RHD simulation of Table~\ref{simus}. }
\label{LDcoeff}
\centering
\renewcommand{\footnoterule}{}  
\begin{tabular}{c|cccc}
\hline \hline
Filter  &$a_{0}$     & $a_{1}$  &  $a_{2}$  \\
\hline
{\sc Mark} 500 nm & 1.000  & $-$0.066 & $-$0.421\\
{\sc Mark} 800 nm & 1.000 & $-$0.041 & $-$0.236\\
{\sc Vinci}               & 1.000 & 0.016 & $-$0.188\\
\hline
\end{tabular}
\end{minipage}
\end{table}

\subsection{Visibility curves}

The synthetic disk images are used to derive interferometric observables. For this purpose, we used the method described in \cite{2009A&A...506.1351C} to calculate the discrete complex Fourier transform $\mathcal{F}$ for each image. The visibility,
$vis$, is defined as the modulus $|\mathcal{F}|$, of the Fourier transform
normalized by the value of the modulus at the origin of the frequency
plane, $|\mathcal{F}_0|$, 
with the phase $\tan\varphi = \Im(\mathcal{F})/\Re(\mathcal{F})$, where $\Im(\mathcal{F})$ and
$\Re(\mathcal{F})$ are the imaginary and real parts of the complex number $\mathcal{F}$,
respectively. 
In relatively broad filters, such as {\sc Vinci}, several spatial frequencies are simultaneously
observed by the interferometer. This effect is called bandwidth smearing. \cite{2003A&A...408..681K, 2003A&A...404.1087K} show that this effect is negligible for squared visibilities larger than 40$\%$ but it is important for spatial frequencies close to the first
minimum of the visibility function. To account for this effect, we computed the squared visibilities as proposed by \cite{2004A&A...413..711W}

\begin{equation}\label{eq_smearing}
\langle vis^2\rangle=\frac{\int_{\lambda_0}^{\lambda_1} vis^2_{\lambda}d\lambda}{\int_{\lambda_0}^{\lambda_1} T^2_\lambda F^2_\lambda d\lambda},
\end{equation}

where $vis^2_{\lambda}$ is the squared visibility at wavelength $\lambda$, $\lambda_0$ and $\lambda_1$ are the filter wavelength limits, $T_\lambda$ is the transmission curve of the filter, and $F_\lambda$ is the flux at wavelength $\lambda$ (see Fig.~\ref{filters}). $vis^2_{\lambda}$ has been computed from the disk images that already include $T_\lambda$ in the calculations.\\
In our case we follow these steps:

\begin{enumerate}
\item we generate a synthetic stellar disk image at different wavelengths of the filters;
\item we compute the visibility curves $vis^2_{\lambda}$ for 36 different cuts through the centers of the stellar disk images;
\item we apply Eq.~(\ref{eq_smearing}) to obtain the averaged squared visibilities.
\end{enumerate}

  Figure~\ref{visibility} (top panels) shows the visibility curves computed with the Eq.~(\ref{eq_smearing}) and for 36 different cuts through the centers of synthetic disk images. This is equivalent to
generating different realizations of the stellar disk images with intensity maps computed for different sets of 
randomly selected snapshots. A theoretical spatial frequency scale expressed in units of inverse solar radii (R$_\odot^{-1}$) is used. The conversion between visibilities expressed in the latter scale and in the more usual \textquotedblleft arcseconds" scale is given by 

\begin{equation}\label{vis1}
vis~[\arcsec]=vis~[{\rm R}_\odot^{-1}]\cdot d~[{\rm pc}]\cdot214.9
\end{equation}

where 214.9  is the astronomical unit expressed in solar radius, and $d$ is the distance of the observed star. The spatial frequency in arcsec$^{-1}$ (i.e, $\nu$) is related to the baseline (i.e., $B$) in meters by 

\begin{equation}\label{vis2}
\nu=\frac{B\cdot4.84813}{\lambda}
\end{equation}

where $\lambda$ is the wavelength in $\mu{\rm m}$.\\
 The first null point of the visibility is mostly sensitive to
the radial extension of the observed object \citep[e.g.][and
 \citeauthor{2010A&A...515A..12C}, \citeyear{2010A&A...515A..12C} for
 an application to RHD simulations]{2001ARA&A..39..353Q}, while the first null point and the second lobe of the visibility curves are sensitive to the limb-darkening \citep{1974MNRAS.167..475H}. Since we want to concentrate on the small scale structure of the surface, the visibility curves of Fig.~\ref{visibility} are plotted longward of the first null point. They show increasing fluctuations with spatial frequencies due to deviations from the circular symmetry relative to uniform disk visibility. This dispersion is clearly stronger in the optical filter at 500 nm and appreciable from longward of the top of third lobe. Moreover, it is also noticeable that the
synthetic visibilities are systematically lower than the uniform disk with a weaker divergence for the {\sc Vinci} filter. This is due to a the non-negligible center-to-limb effect visible in Figs.~\ref{images} and \ref{profiles}. 

The bottom panel of Fig.~\ref{visibility} displays the one $\sigma$ visibility fluctuations, $F$, with respect to the average value $\overline{{vis}}$ ($F=\sigma/\overline{{vis}}$). The dispersion increases with spatial frequency. This is due to the small scale structure on the model
stellar disk \citep[see e.g.][]{2010A&A...515A..12C}. The dispersion is stronger in the case of the {\sc Mark}~III 500 nm filter with respect to the redder filters. \\
Figure~\ref{profiles} shows small differences between the averaged intensity profiles of the {\sc Mark}~III 500 nm filter with and without considering the spectral lines in our calculations. Therefore, we computed visibility curves in both cases and found that the visibility fluctuations are indistinguishable (Fig.~\ref{visibility}, bottom panel). While the molecular absorption can cause a strong difference in stellar surface appearance (and consequentially also on the visibility curves) in the case of cool evolved stars \cite[e.g., the contribution of H$_2$O to the radius measurement for red supergiants stars, ][]{2010A&A...515A..12C}, this is not the case for Procyon where the atomic lines are not strong enough to cause an appreciable effect and also the atmosphere is very compact.\\
The bottom panel of Fig.~\ref{visibility} also shows that, on the top of the second lobe ($\sim0.4$~R$^{-1}_\odot$), the fluctuations are of the order of $\sim0.5\%$ of the average value for {\sc Mark}~III 500 nm filter  and $\sim0.1\%$ for {\sc Mark}~III 800 nm and {\sc Vinci} filters. From the top of the third lobe ($\sim0.6$~R$^{-1}_\odot$) on, the fluctuations are $\sim2\%$ of the average value in the optical region, which is larger of the instrumental error of VEGA on CHARA \citep[$\sim$1$\%$, ][]{2009A&A...508.1073M}. \\
It must be noted that our method of constructing realizations of stellar disk images inevitably introduces some discontinuities between neighbouring tiles by randomly selecting temporal snapshots and by cropping intensity maps at high latitudes and longitudes. However, \cite{2010A&A...524A..93C} proved that for
the signal artificially introduced in to the visibility curves is largely weaker than the signal due to the inhomogeneities of the stellar surface.

\section{Fundamental parameters}

\subsection{Multiwavelength angular diameter fits in the {\sc Mark}~III and {\sc Vinci} filters}\label{multiradius}

We used the data from the {\sc Mark}~III interferometer at 500 and 800 nm \citep{1991AJ....101.2207M} plus additional measurements at 500 nm reported in \cite{2005ApJ...633..424A}. It has to be noted that \cite{1991AJ....101.2207M} do not provide a calibrator list even thought they claim that the calibrators are all smaller that the science star. The {\sc Mark}~III 500 nm filter has points on the first and the second lobe while the {\sc Mark}~III 800 nm has limited baselines in the first lobe. For the {\sc Mark}~III 500 nm filter, we used only the data with baselines larger than 20 meters because at shorter baselines the error bars are $>10\%$ of the squared visibility.
We also used {\sc Vinci} data from the Very Large Telescope Interferometer \citep{2004A&A...425.1161K,2005ApJ...633..424A}. The wavelength of these observations is around 2.2 $\mu$m and the baselines go up to 66 m, significantly lower in the first lobe of the Procyon visibility curve.

\begin{figure*}
   \centering
   \begin{tabular}{ccc}
   \includegraphics[width=0.33\hsize]{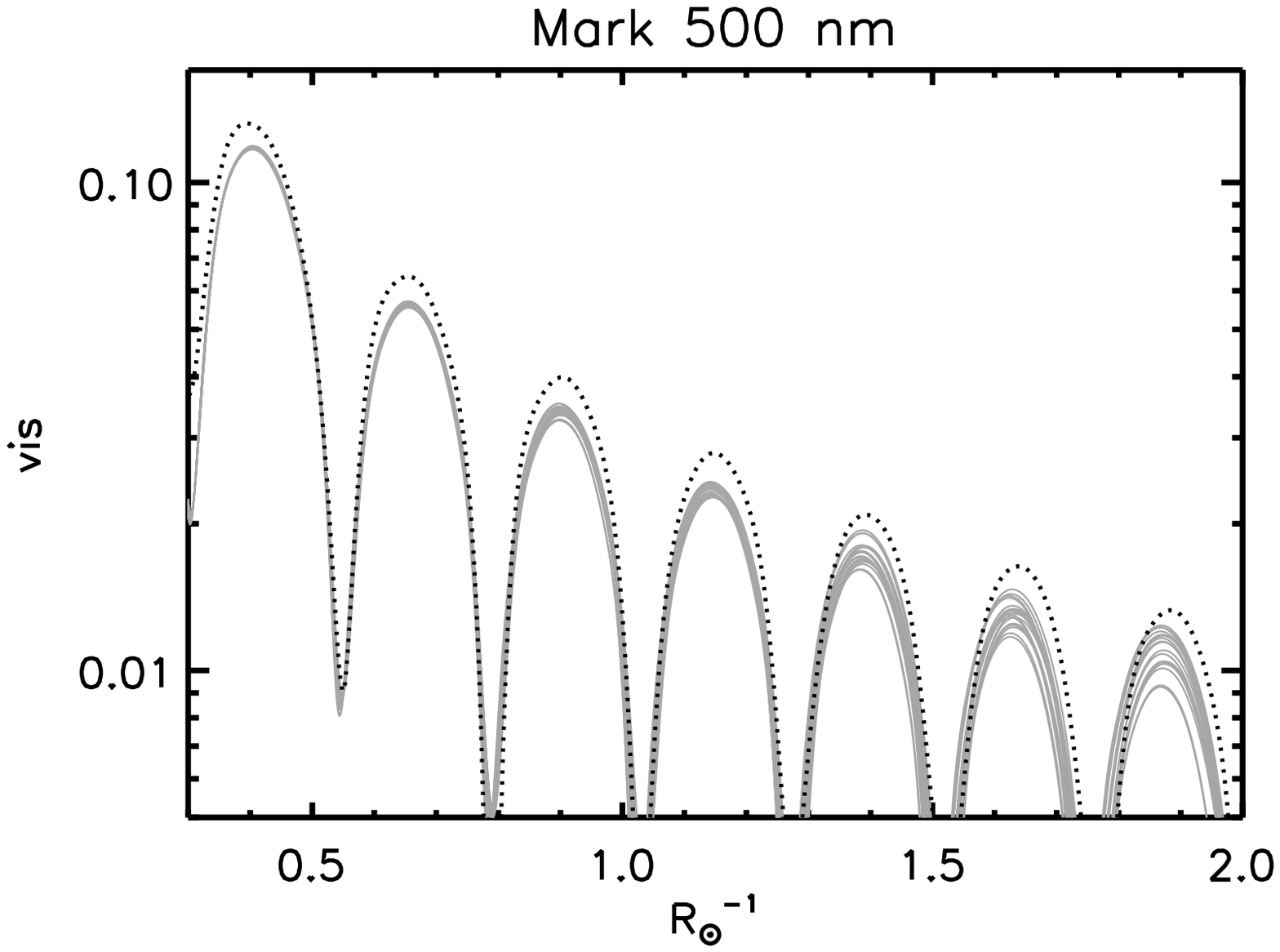}
       \includegraphics[width=0.33\hsize]{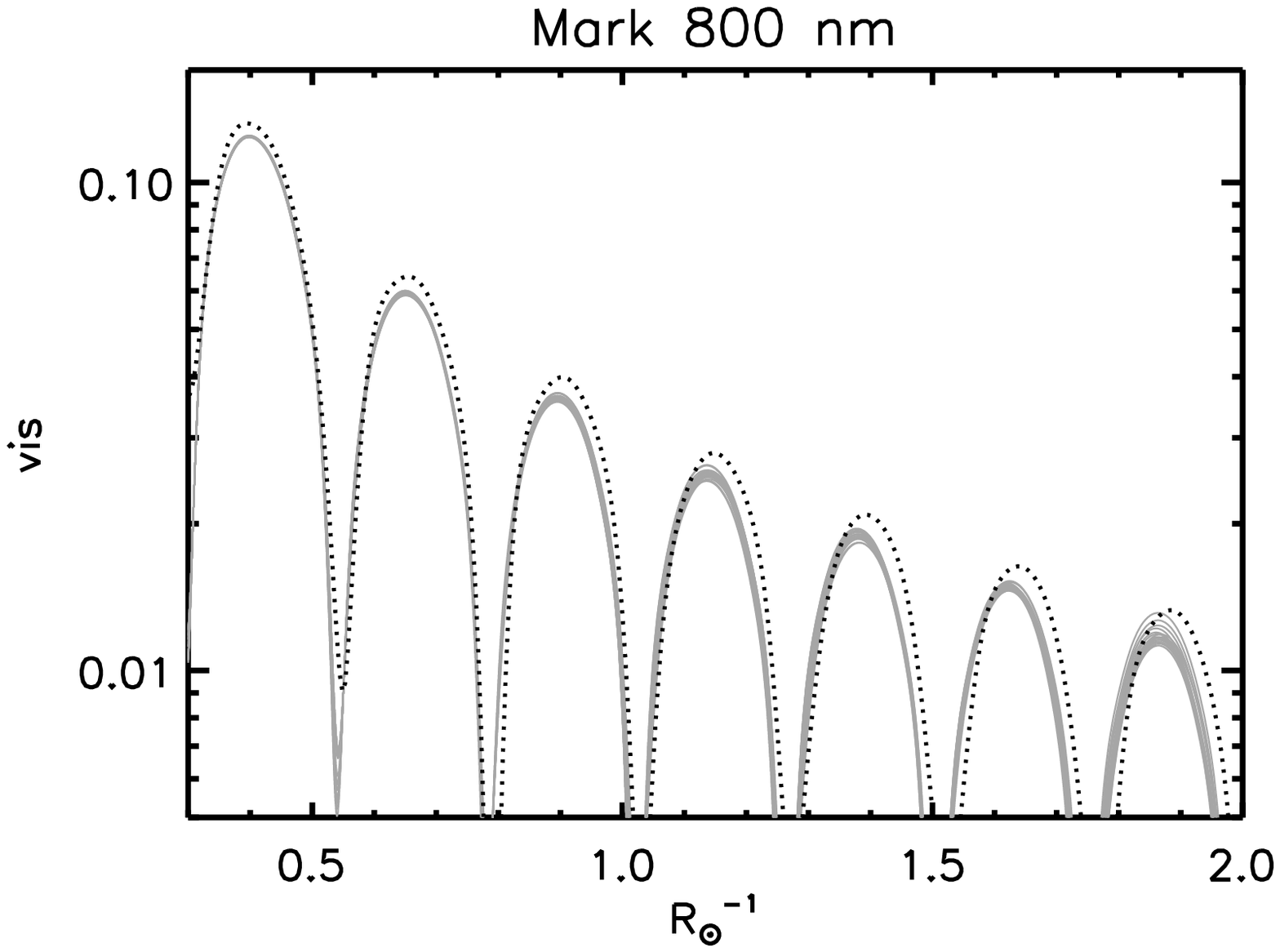}
           \includegraphics[width=0.33\hsize]{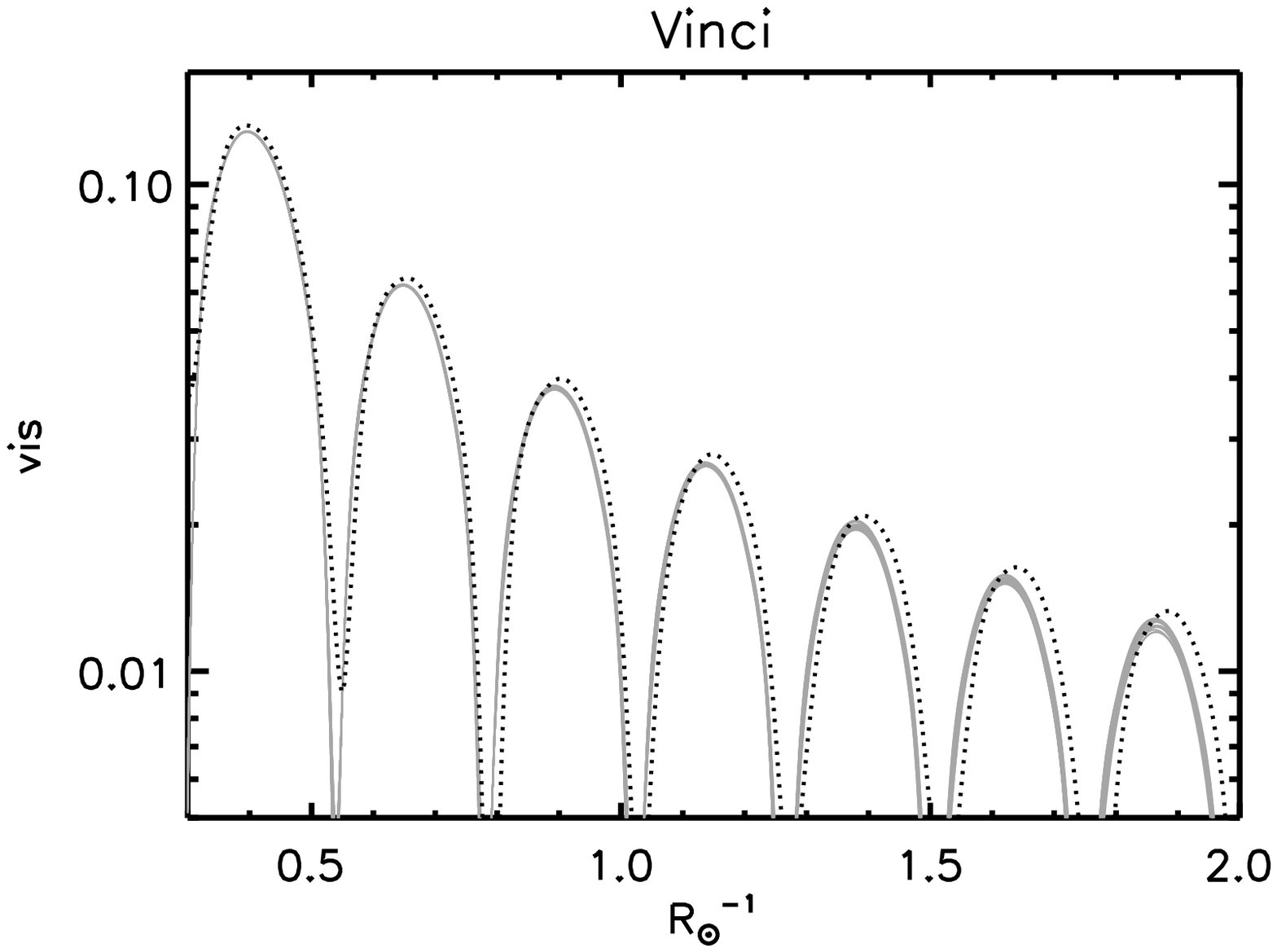}\\
             \includegraphics[width=0.4\hsize]{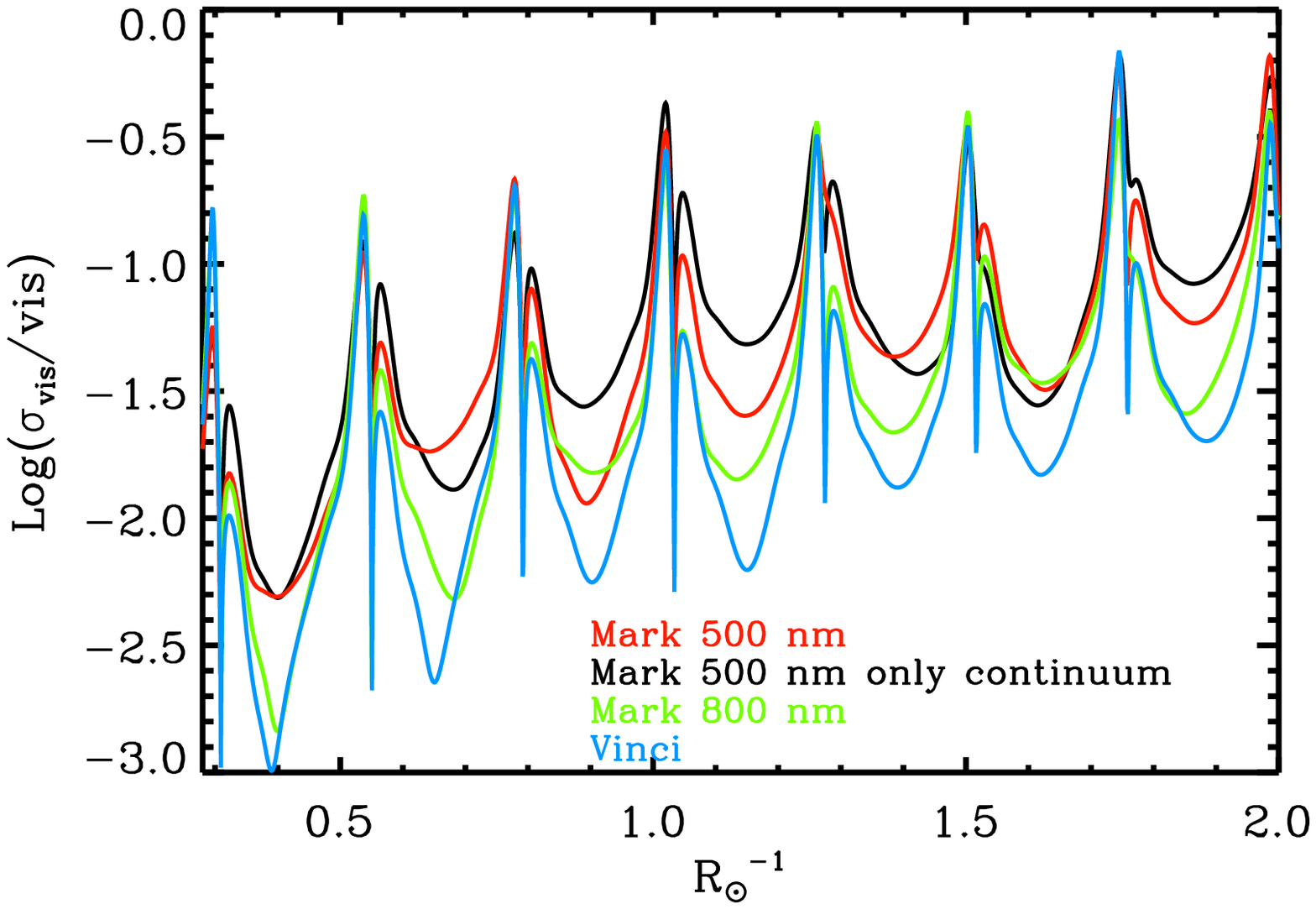}
        \end{tabular}
      \caption{\emph{Top panels}: visibility curves ($vis$) from the Procyon simulation in the filters computed in this work. The visibilities are computed for 36 different azimuth angles 5$^\circ$ apart (thin gray lines). The dotted line is an uniform disk scaled to match the same radius, i.e. the first null in visibility curve. The visibilities are displayed only longward of the first null visibility point. A logarithmic scale is used on y-axis. \emph{Bottom panel}: visibility fluctuations ($\sigma/\overline{{vis}}$) with respect to the visibility average value ($\overline{{vis}}$) as a function of spatial frequencies for all the considered filters.}
        \label{visibility}
   \end{figure*}

We used two independent methods to determine the angular diameters: 

\begin{enumerate}
\item the synthetic visibilities of Fig.~\ref{visibility} obtained from the spherical tiling method described in this work and scaled to absolute model dimensions (using Eq.~(\ref{vis1}) and (\ref{vis2})) to match the interferometric observations;
\item the method described in \cite{2005ApJ...633..424A,2006A&A...446..635B,2011A&A...534L...3B}, that computes the normalized fringe visibilities using the van Cittert-Zernike theorem.
\end{enumerate}

\begin{figure}
   \centering
    \begin{tabular}{ccc}
   \includegraphics[width=0.95\hsize]{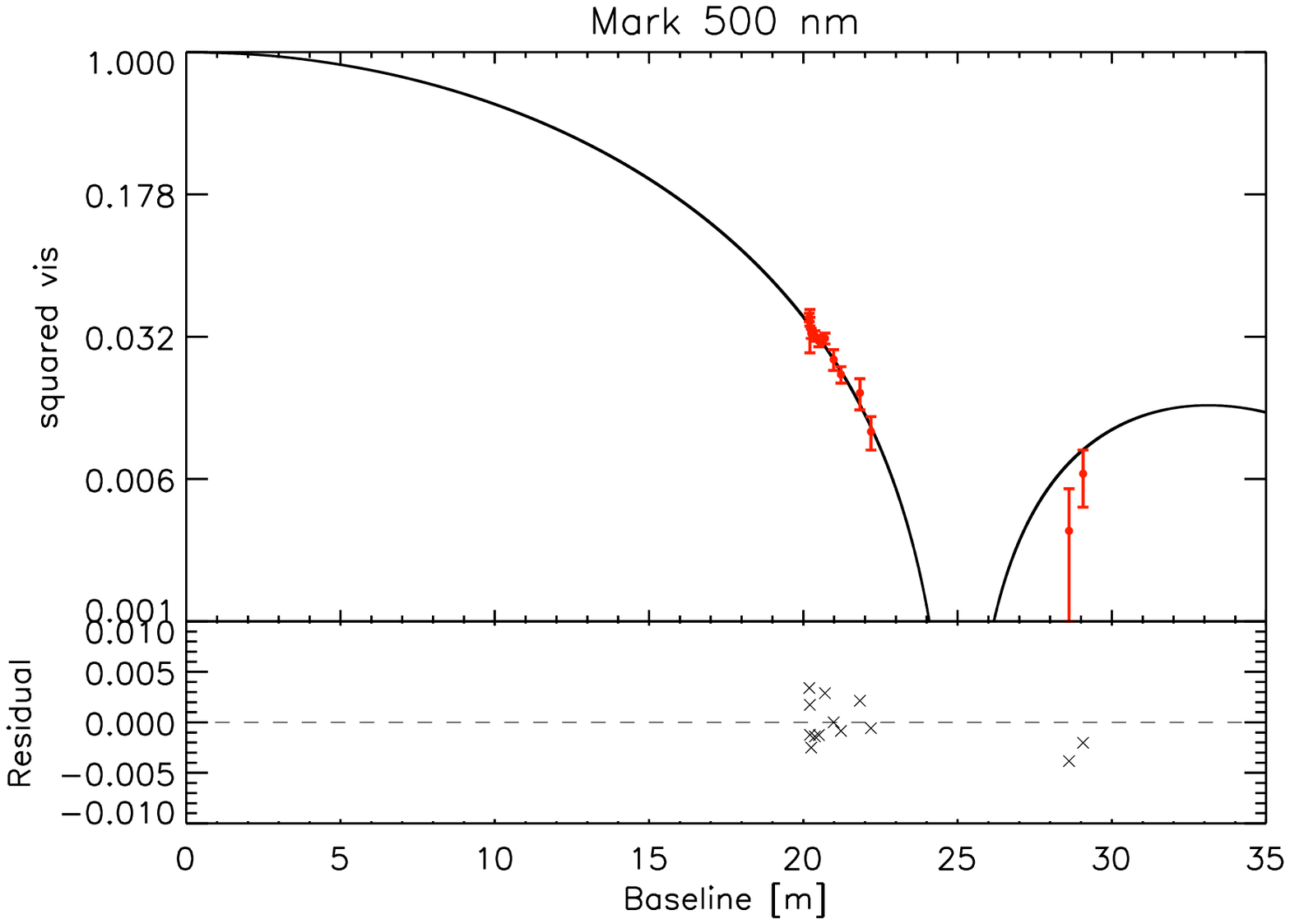}\\
   \includegraphics[width=0.95\hsize]{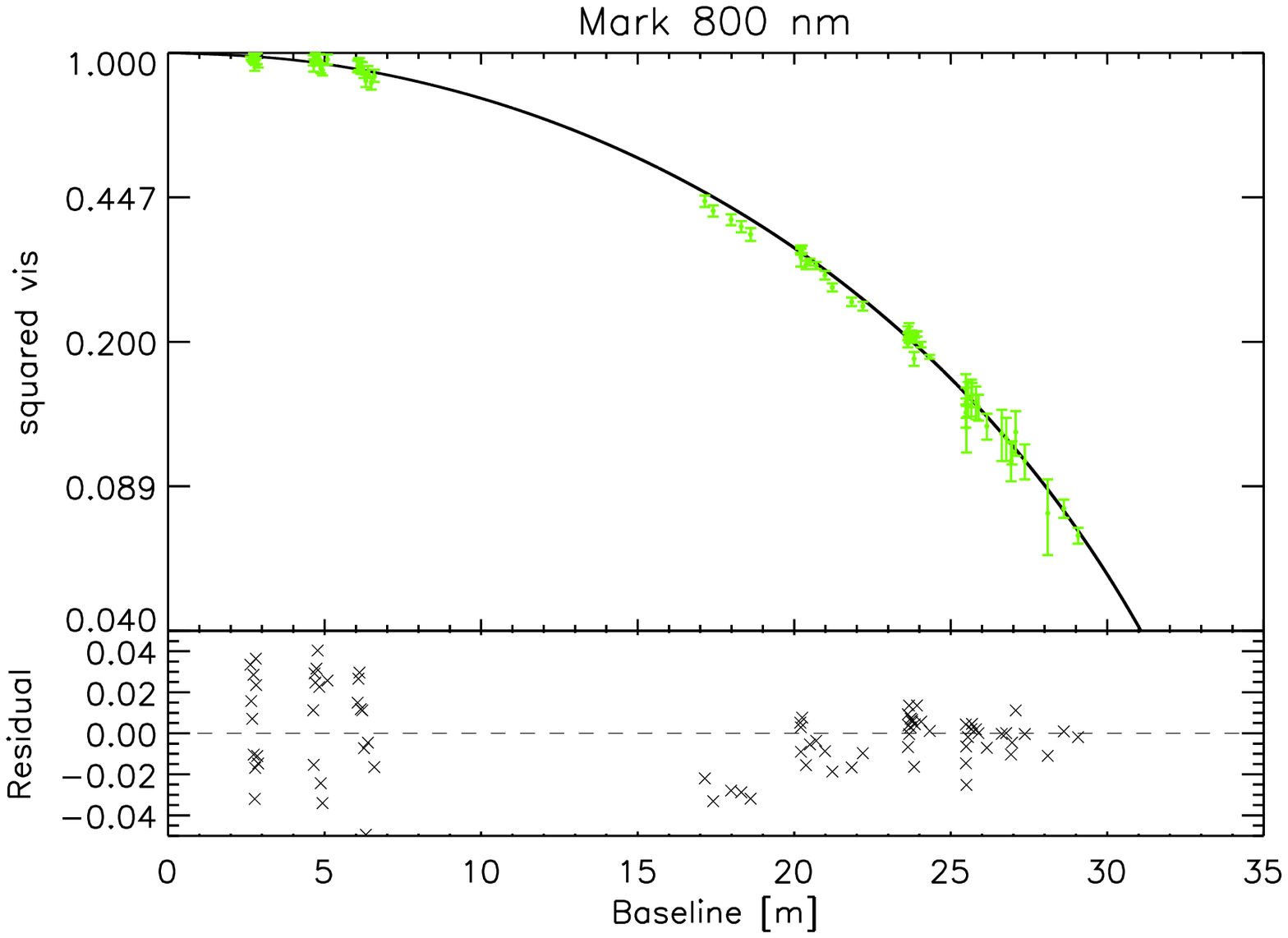}\\
   \includegraphics[width=0.95\hsize]{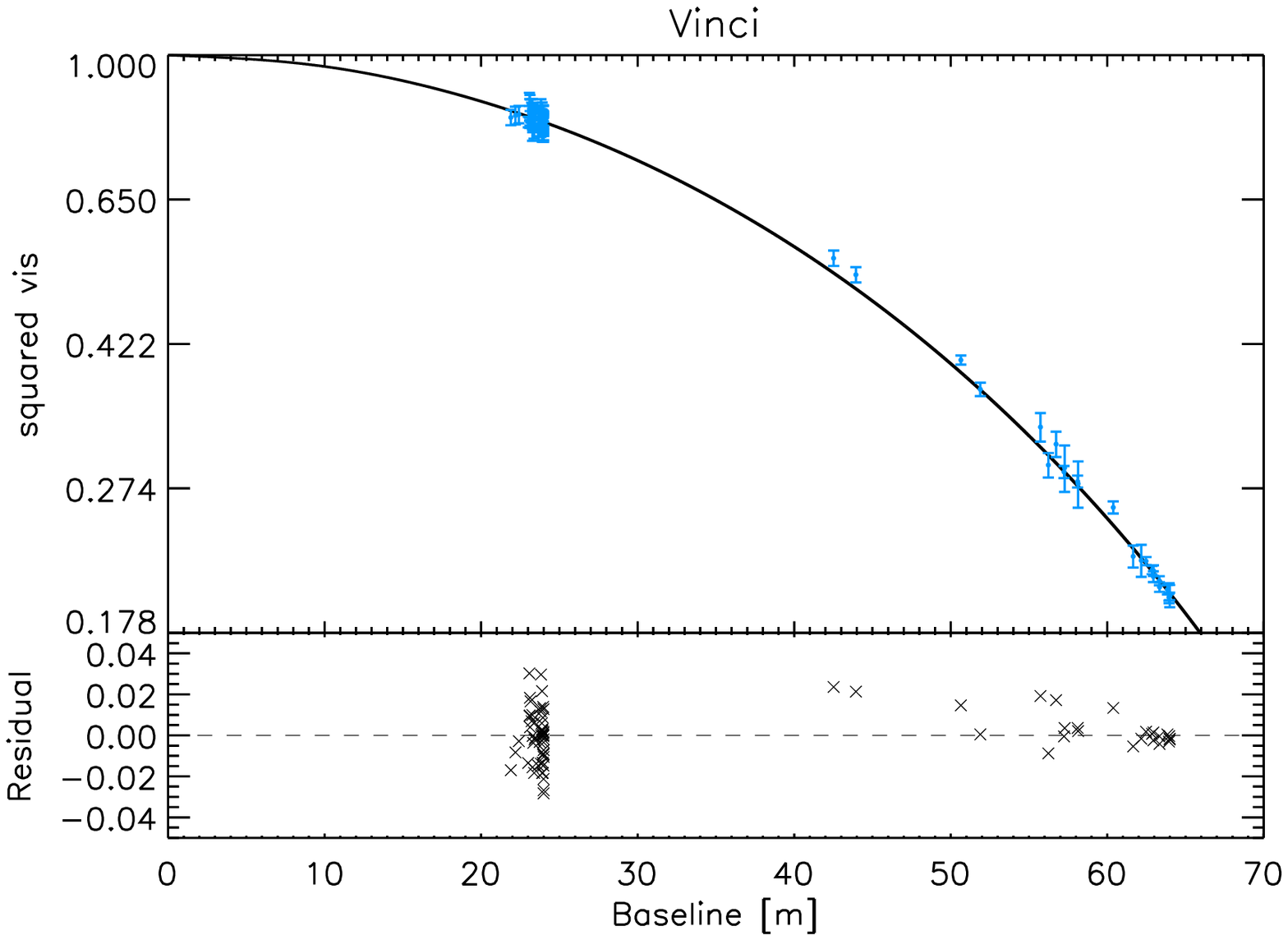}
    \end{tabular}
    \caption{Best matching synthetic visibility from Fig.~\ref{visibility} for the {\sc Mark} 500, {\sc Mark} 800, and {\sc Vinci} data. The stellar disk maps in each filter have been scaled to match the radius of the star. For the {\sc Mark} 500 nm filter, we used only the data with baselines larger than 20 meters.
    }
              \label{observing_fit}%
    \end{figure}

Method (1) is a the direct consequence of the Fourier Transform of the synthetic disk images (Fig.~\ref{images}): the visibilities vary for different cuts  through the centers of synthetic disk images (i.e., the position of the first null changes its position). Method (2) is based on the integration of the spatial average intensity profile. For both methods, we used a Levenberg-Marquardt least-square minimization.\\
Table~\ref{diameters} reports the best-fit angular diameters. While for the uniform disk there is a significantly wide range of values among the different filters, the angular diameters from the RHD simulation are closer and overlap with the uncertainties. The two independent methods used in Table~\ref{diameters} show a clear tendency of the optical diameters to appear smaller than in the infrared ($\frac{\theta_{\rm{{\sc Mark}~III}}}{\theta_{\rm{{\sc Vinci}}}} \sim 0.99$). The possible explanation for the different radius between {\sc Mark}~III and {\sc Vinci} are: (i) the RHD model used is not fully appropriate to the observations, for instance the center-to-limb variation may not be correct because, for instance, the points on the second lobe (sensitive to the center-to-limb variation) of {\sc Mark} 500 nm filter do not match very well (Fig.~\ref{observing_fit}, top panel); (ii) the data of {\sc Mark}~III may present problems (e.g., calibrator or systematics). \\
In addition to this, we also conclude that the two methods give consistent results and can be used without distinction to perform angular diameter fits at least for dwarf stars. If fact, in the case of cool evolved stars with low surface gravity ($\logg<2.0$), the surface asymmetry may strongly impact the shape of the star and the radius depends on orientation of the projected baseline \citep{2009A&A...506.1351C}.

The angular diameters found for the {\sc Vinci} filter (Table~\ref{diameters}) are in fairly good agreement with \cite{2005ApJ...633..424A} who found 5.403 $ \pm $ 0.006 mas. However, $\theta_{\rm{{\sc Vinci}}}$ is $\sim2\%$ smaller than: (i) 5.448 $ \pm $ 0.03 mas \citep{2004A&A...413..251K}, based on the fit of the {\sc Vinci} with baseline points of Fig.~\ref{observing_fit} lower than 22 meters; (ii) 5.48 $ \pm $ 0.05 mas \citep{2002ApJ...567..544A}, obtained with Eq.~\ref{eq_seismic} from the linear radius of 2.071 $ \pm $ 0.020 $R_\odot$; (iii) 5.50 $ \pm $ 0.17 mas \citep{1991AJ....101.2207M}. Our $\theta_{\rm{{\sc Vinci}}}$ is also close to 5.326 $ \pm $ 0.068 mas found by \cite{2010A&A...512A..54C} with the infrared flux method.

Transforming the {\sc Vinci} angular diameter to linear radius (Eq.~\ref{eq_seismic}), we found 2.019 $R_\odot$. This value is compared to what we chose as a reference in Table~\ref{simus}: 2.055 $R_\odot$. The difference is 0.023 $R_\odot$, which is 16 Mm (i.e., 0.7 times the numerical box of RHD simulation, 22 mM, Table~\ref{simus}, 4th column). We checked that this has a negligible impact on the observables found in this work using the spherical tiling method described in Section~\ref{tilingsect}.\\
The astrometric mass \citep[$M=1.430 \pm 0.034\,M_{\odot}$][used in Section~\ref{asterosection}]{gatewood06} combined with our interferometric diameter leads to a new gravity $\logg = 4.01 \pm 0.03$ [cm/s$^2$], which is larger by $0.05$ dex than the value derived in \citet{2002ApJ...567..544A}. The contribution of the revised Hipparcos parallax is $\sim$0.01 dex in the $\logg$ change.

\begin{table}
\begin{minipage}[t]{\columnwidth}
\caption{Best angular diameters, in mas, for RHD simulation with Method (1) ($\theta_{\rm{Method1}}$) and Method (2) ($\theta_{\rm{Method2}}$) described in Section~\ref{multiradius}; and for the uniform disk. The error is one $\sigma$ with respect to the average value.}
\label{diameters}
\centering
\renewcommand{\footnoterule}{}  
\begin{tabular}{c|cccc}
\hline \hline
Filter  & $\theta_{\rm{Method1}}$ & $\theta_{\rm{Method2}}$  & $\theta_{\rm{UD}}$ \\
\hline
{\sc Mark} 500 nm & 5.313 $ \pm $ 0.03 & 5.324 $ \pm $ 0.03 &  5.012 $ \pm $ 0.03 \\
{\sc Mark} 800 nm & 5.375 $ \pm $ 0.06 & 5.370 $ \pm $ 0.06 & 5.208 $ \pm $ 0.06 \\ 
{\sc Vinci}               & 5.382\footnote{Since the resulting diameters are very similar, we adopted an average angular diameter of 5.390 for all the calculations in the article}    $ \pm $ 0.03 & 5.397$^a$ $ \pm $ 0.03 & 5.326 $ \pm $ 0.03 \\ 
\hline
\end{tabular}
\end{minipage}
\end{table}

\subsection{The effective temperature} \label{sec:teff}

The effective temperature is determined from the bolometric flux $\fbol$ and the angular diameter $\theta$ in the {\sc Vinci} filter (average between the two methods in Table~\ref{diameters}, $\theta$ = 5.390 mas) by the relation

\begin{equation}\label{teff}
\teff =  \left ( \frac{4 \fbol}{\sigma_{\rm{st}} \theta^2}  \right )^{0.25},
\end{equation}

where $\sigma_{\rm{st}}$ stands for the Stefan-Boltzmann constant. Our derived diameter being smaller by $\sim2$\% than \citet{1991AJ....101.2207M} or \citet{2004A&A...413..251K}, the derived effective temperature has to be larger by $\sim1$\%  compared to \citet{2002ApJ...567..544A} or \citet{2005ApJ...633..424A}, who used the previous mentioned references.  There are several sources for the bolometric flux leading to slightly different $\teff$. The values  of   $\fbol= (18.20 \pm 0.43)\, 10^{-6}\,{\rm \,erg\,cm^{-2}\,s^{-1}}$  \citep{fuhrmann97}  and $\fbol = (17.82 \pm 0.89)\, 10^{-6}\,{\rm \,erg\,cm^{-2}\,s^{-1}}$ \citep{2005ApJ...633..424A} lead to $\teffhydro =  6591 \pm 43 $ K and $6556\pm 84$ K, respectively. 

Our new 3D $\teffhydro$ returns a value closer to $\teffir= 6621 \pm 80$ K obtained by \citet[][and Casagrande private communication]{2010A&A...512A..54C} than the old derived value of $6516 \pm 87$ K   \citep{2005ApJ...633..424A}.\\
 The influence of the uncertainties in the selected fundamental parameters ($\teff$,$\logg$,$\feh$) of our RHD model atmosphere has a negligible impact on the limb-darkening and  therefore on the derived angular diameter and $\teff$. This  was tested for HD~49933 \citep{2011A&A...534L...3B}, a star similar to Procyon.
   
\subsection{Asteroseismic independent determination of the radius}\label{asterosection}

The radius of the star can be derived from its oscillation spectrum. The frequency  $\numax$
 of the maximum in the power spectrum  is generally assumed to scale with the acoustic cut-off frequency of the star \citep[e.g.][]{brown91,kjeldsen11}, therefore $\numax \approx g/\sqrt\teff$. Then, it is  straightforward to derive the radius 
 \begin{equation}\label{eq:scaling}
 \frac{R}{R_{\odot}} =  \left ( \frac{M}{M_{\odot}} \frac{\numaxsol}{\numax} \right )^{0.5} \left ( \frac{\teff}{\teffsol} \right )^{-0.25}
 \end{equation}
The validity of such scaling relation has been verified on large asteroseismic surveys \citep[e.g.][]{bedding03,verner11}. The value of $\numax$ derived from photometry  is accurately determined,  $ 1014 \pm 10\,\muhz$ \citep{arentoft08}. The solar value of $\numax$ is taken from \citet{belkacem11}.
We emphasize  that the dependence on $\teff$  in Eq. \ref{eq:scaling}  is  weak, therefore the derived radius is not very sensitive to the selected value of the effective temperature. We use the value of $\teff=6591 \pm 43$K derived in Section \ref{sec:teff}, since it is closer to the infrared flux method determination. 
Since Procyon is a binary star, the mass can by determined by the astrometric orbital elements and the third Kepler's law. However, the derived value is subject to debate. Indeed, \citet{girard00} found a mass of $M=1.497 \pm 0.037\,\sunmass$, whereas \citet{gatewood06} found $M=1.430\pm 0.034 \sunmass$. As discussed in the introduction, we prefer to keep the value of \citeauthor{gatewood06} since stellar evolution models that use this mass, are in agreement with the age of the white dwarf companion.
Using these values,  we found a radius of ${\rm R=2.023 \pm 0.026\,R_{\odot}}$.  We can translate this radius into angular diameter using  the relation
\begin{equation}\label{eq_seismic}
\theta_{\rm seismic} =  2\, \left(R/R_{\rm\odot}\right)\, \pi_p \tan (\theta_{\rm\odot}/2) = 5.36 \pm 0.07 \,{\rm mas},
\end{equation}
with the solar angular radius $\theta_{\rm\odot}/2 = 959.64 \pm 0.02$ arcsecs \citep{chollet99} and the parallax $\pi_p = 284.56 \pm 1.26$ mas  \citep{vanleeuwen07}.  This radius agrees well with our interferometric value within error bars.  

\section{Spectrophotometry and Photometry}

\begin{figure}
   \centering
    \includegraphics[width=0.98\hsize]{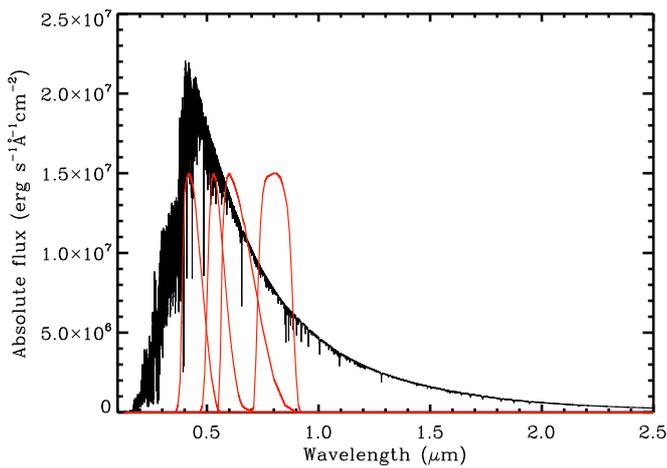}
         \caption{Spatially and temporally average synthetic spectral energy distribution from the UV to the near infrared. Red curves correspond to \cite{1990PASP..102.1181B} for the filters $BVRI$.}
        \label{sed}
   \end{figure}

The radiative transfer code {\sc Optim3D} includes all the up-to-date molecular and atomic line opacities. This allows for very realistic spectral synthetic RHD simulations for wavelengths from the ultraviolet to the far infrared. It is then possible to compute realistic synthetic colors and compare them to the observations. Figure~\ref{sed} shows the spectral energy distribution (SED) computed along rays of four $\mu$-angles [0.88, 0.65, 0.55, 0.34] and four $\phi$-angles [$0^{\circ}$,
$90^{\circ}$, $180^{\circ}$, $270^{\circ}$], and after a temporal average over all selected
snapshots (like in Fig.~\ref{filters}). The shape of SED reflects the mean thermal gradient of the simulations. 

We used the prescriptions by \cite{1990PASP..102.1181B} to compute the color indexes for the filters $BVRI$. Table~\ref{tablecolors} displays the comparison of the synthetic colors with the observed ones. The difference is very small.

\begin{table}
\centering
\caption{Photometric colors for RHD simulation of Table~\ref{simus} and for the corresponding observations of \cite{1990A&AS...83..357B}}
\label{tablecolors}      
\renewcommand{\footnoterule}{} 
\begin{tabular}{c c c c c }        
\hline\hline                 
  		     & $B-V$   &  $V-R$  & $R-I$   &  $V-I$   \\ 
\hline
RHD simulation    &  0.419 & 0.255 &  0.252 & 0.507      \\  
Observation &   0.420    &  0.245 &   0.245 &  0.490  \\

\hline\hline                          
\end{tabular}
\end{table}

We used the absolute spectrophotometry measurements at ultraviolet and visual wavelengths collected by \cite{2005ApJ...633..424A} to compare with the synthetic SED. The data come from: (i) the Goddard High Resolution Spectrograph (GHRS) data sets Z2VS0105P ( PI A. Boesgaard), Z17X020CT, Z17X020AT, Z17X0208T (PI J. Linsky) from 136 and 160 nm; (ii) the International Ultraviolet Explorer (IUE) \cite{1999A&AS..139..183R} from 170 to 306 nm; (iii) the Hubble Space Telescope imaging Spectrograph (STIS) from 220 to 410 nm; (iv) and the visual and near-infrared wavelength data from \cite{1992A&AS...92....1G}.
The data from the GHRS are far superior to any other measurements below 160 nm because the continuum drops by more than a factor of 100 here, too much for the limited dynamic range of IUE. The flux was estimated by computing the mean flux between the emission lines in each spectrum incorporating the flux uncertainties provided with each calibrated data set \citep{2005ApJ...633..424A}.\\

\begin{figure*}
   \centering
   \begin{tabular}{cc}
   \includegraphics[width=0.5\hsize]{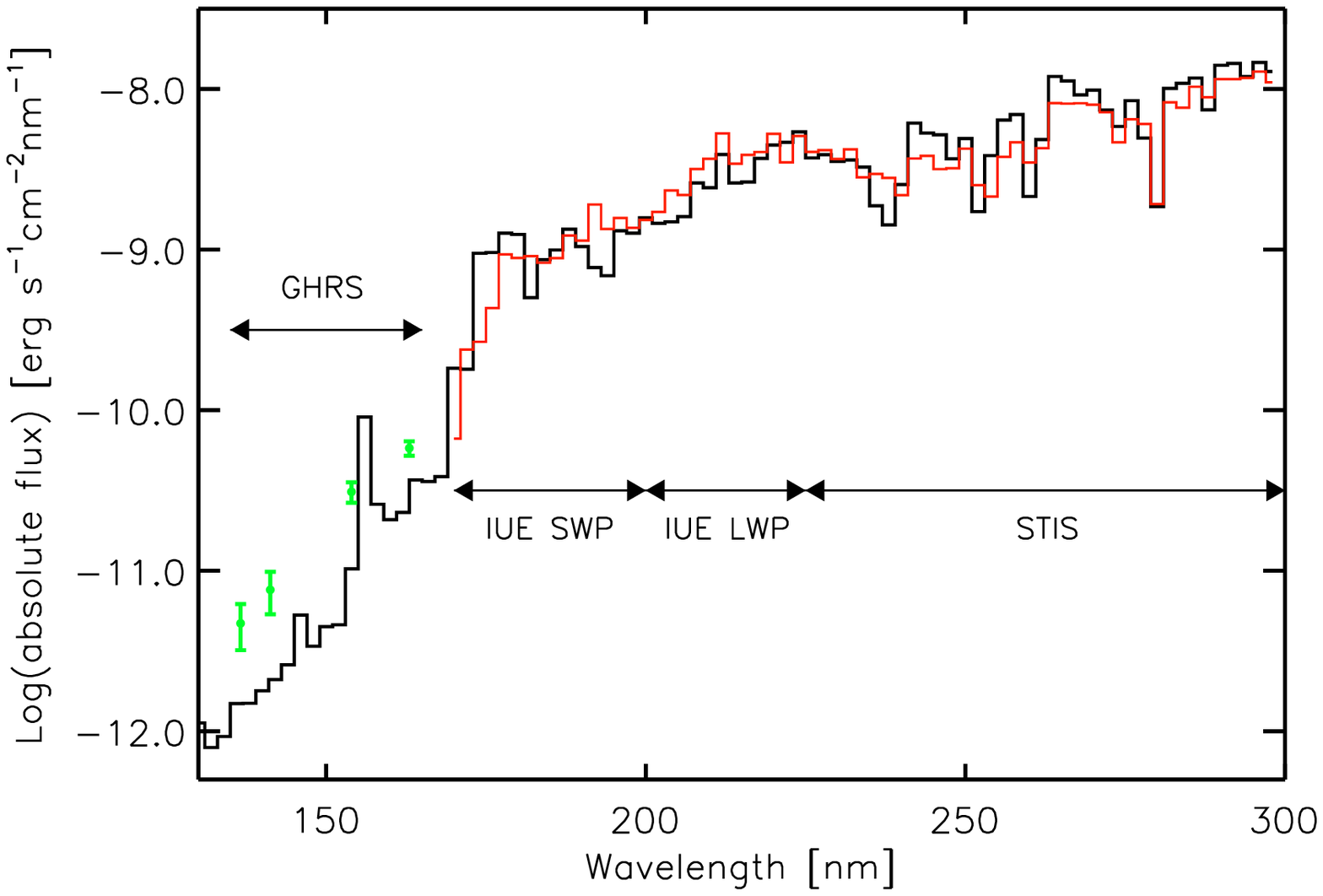}
       \includegraphics[width=0.5\hsize]{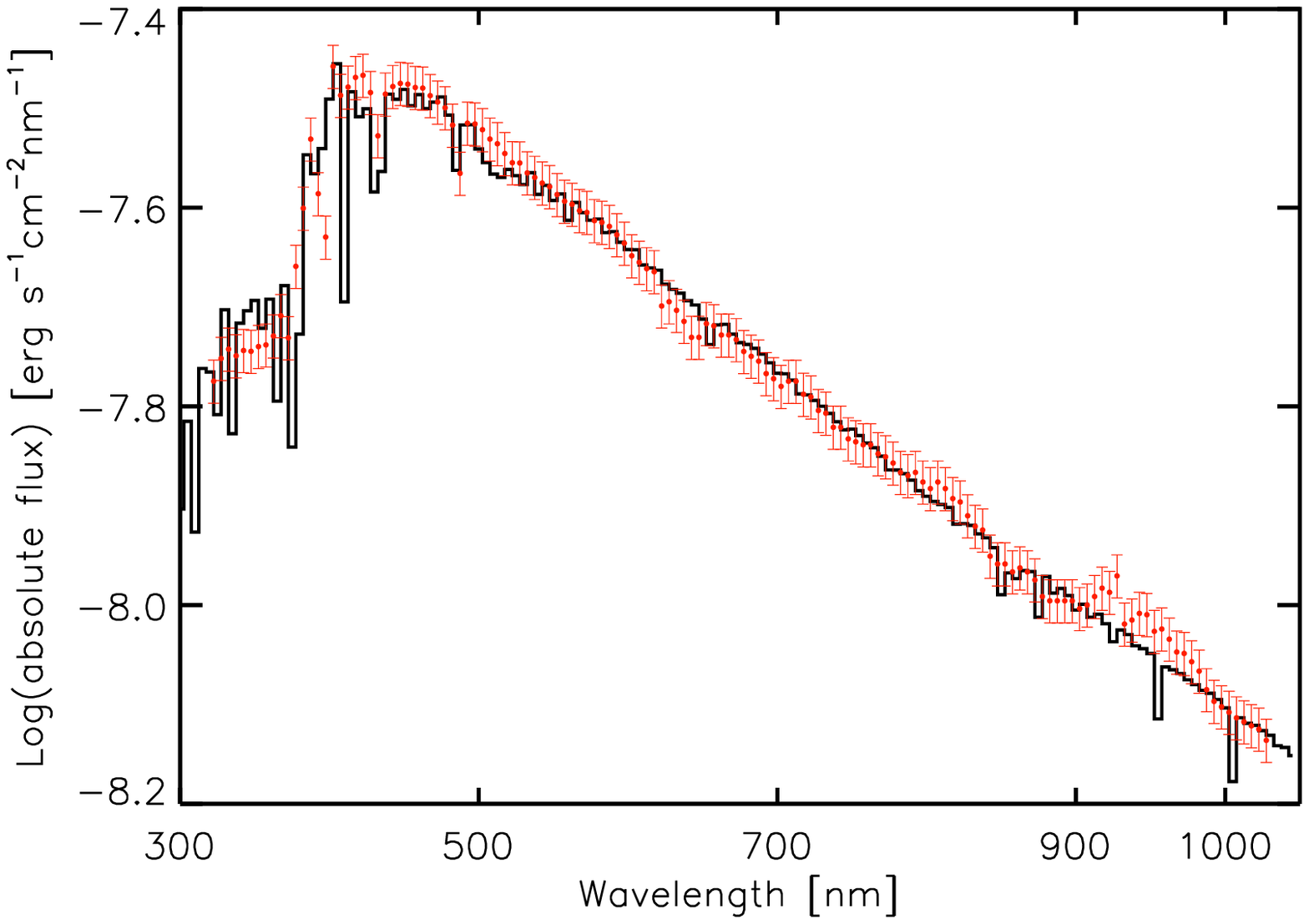}\\
          \includegraphics[width=0.5\hsize]{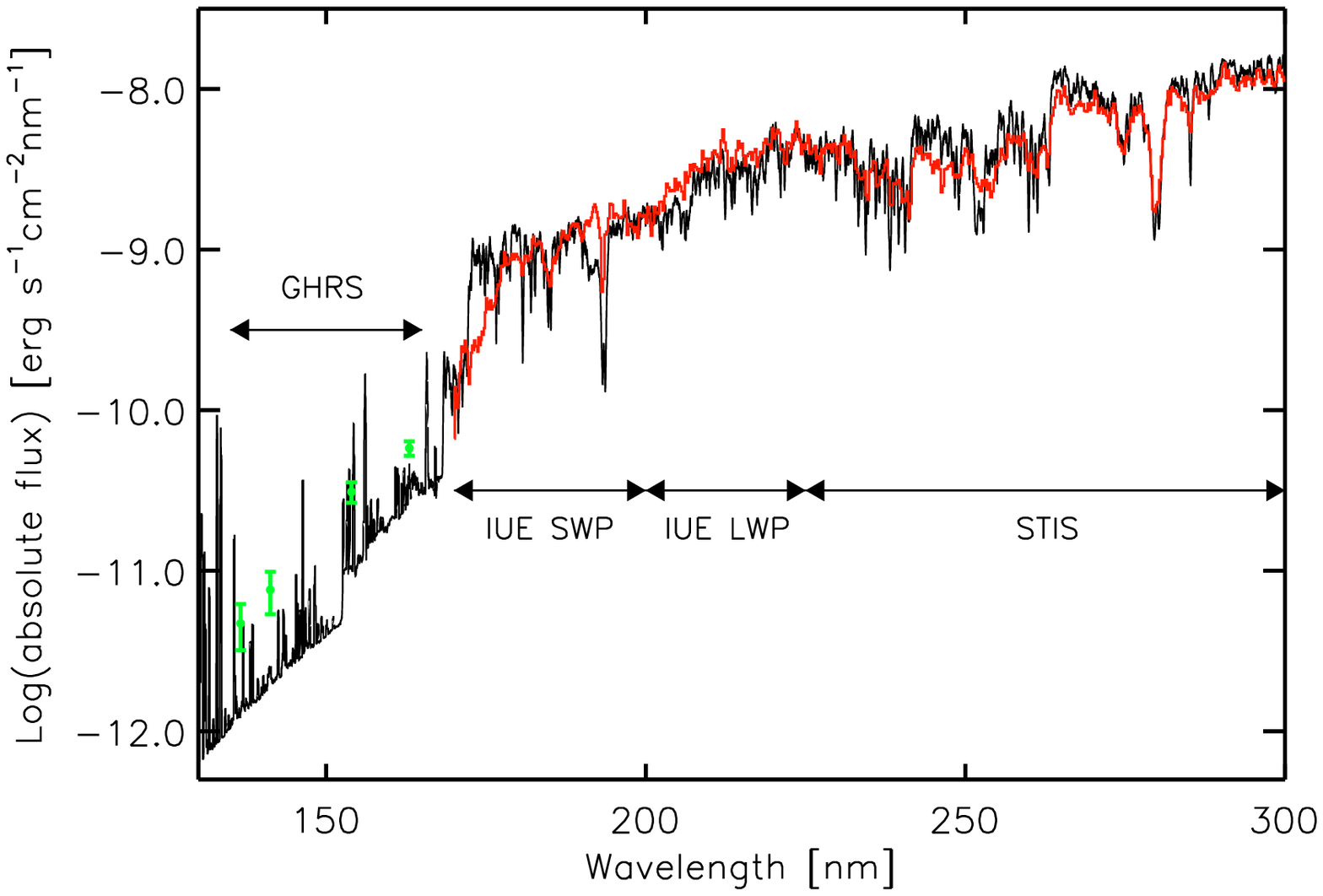}
       \includegraphics[width=0.5\hsize]{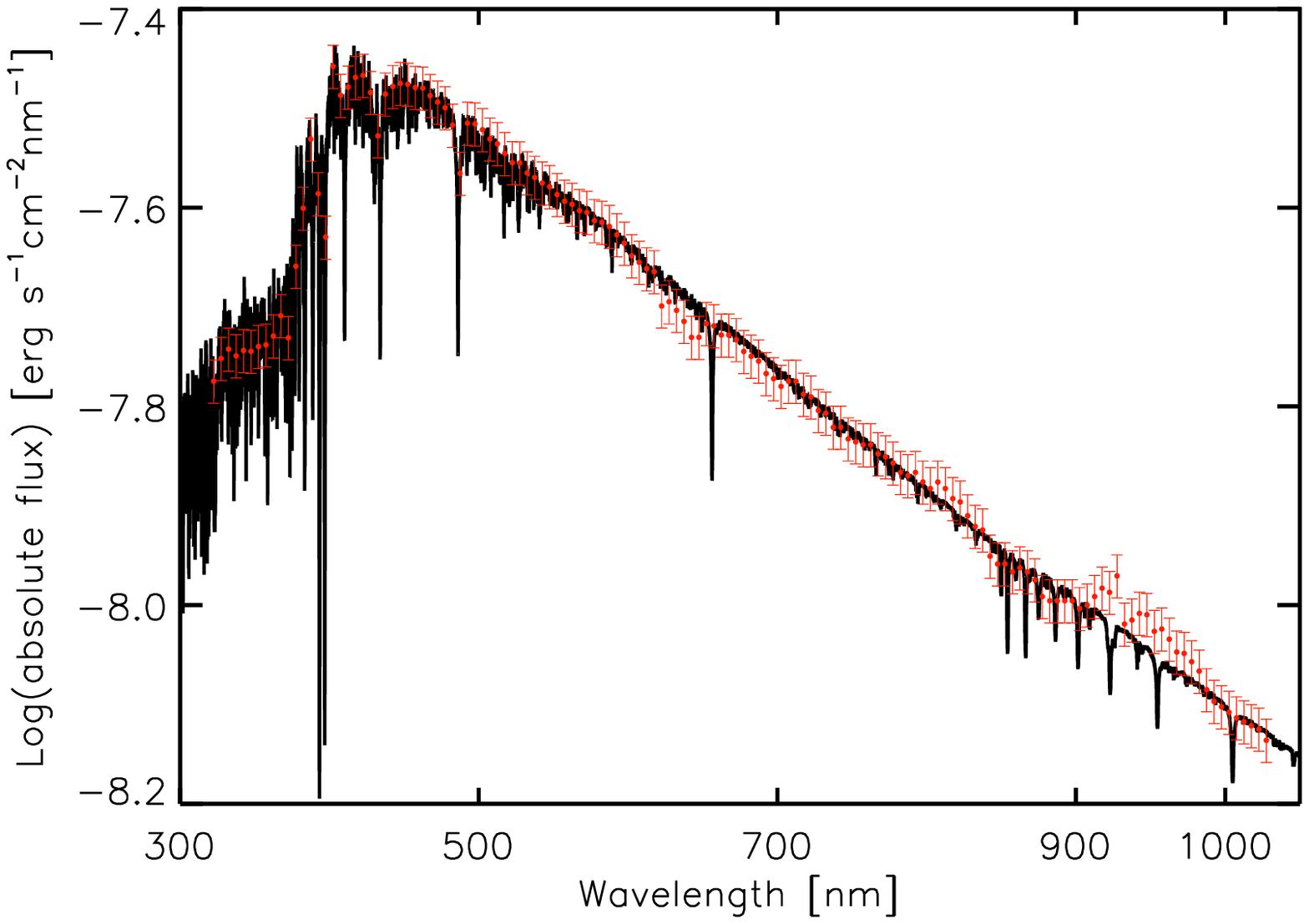}
        \end{tabular}
      \caption{Comparisons between synthetic SED (black) and spectrophotometric measurements at ultraviolet and visual wavelengths (green dots with error bars in left column plots and red line). \emph{Top row:} the models and data are binned to 2 nm resolution in the UV for clarity and to 5 nm in the visual to match the resolution of the observed spectrophotometry of \cite{1992A&AS...92....1G}. \emph{Bottom row:} same as above with the synthetic SED and the observations at higher spectral resolution. The synthetic SEDs are scaled in absolute flux using an angular diameter of 5.390 mas found from the fit of {\sc {\sc Vinci}} filter observations.}
        \label{photometry}
   \end{figure*}

 Figure~\ref{photometry} (top row) shows that our synthetic SED matches fairly well the observations from the UV to the near-IR region. This result can be compared to the already good agreement found with composite RHD model done with CO$^5$BOLD code in \cite{2005ApJ...633..424A}. Moreover, the bottom row of the Figure shows that at higher spectral resolution, the agreement is even more remarkable. RHD models match also the observations between 136 and 160 nm, that are supposed to form at depths beneath its chromosphere \citep{2005ApJ...633..424A} and thus being affected by the convective-related surface structures. It should also be noted that the ultraviolet SED longward of 160 nm may be impacted by non-LTE treatment of iron-group elements. \cite{2005ApJ...618..926S} found that non-LTE models for the Sun have up to 20 $\%$ more near-UV flux relative to LTE models. It is not possible to determine if this difference is also present in Procyon because either the RHD model calculation and the post-processing calculations have been done with LTE approximation. The ultraviolet SED may also be affected by scattering at these wavelengths. This effect is also not included in our calculations. However, \cite{2010A&A...517A..49H} demonstrated that, in RHD simulations, the scattering does not have a significant impact on the photospheric temperature structure in the line forming region for a main sequence star.\\
We conclude that the mean thermal gradient of the simulation, reflected by the spectral energy distribution, is in very good agreement with Procyon.

\section{Closure phases and perspectives for hot Jupiter detection}

Interferometry has the potential for direct detection and characterization of extrasolar planets. It has been claimed that differential interferometry could be used to obtain spectroscopic information, planetary mass, and orbit inclination of extrasolar planets around nearby stars: \cite{2000SPIE.4006..269S,2000SPIE.4006..407L,2004SPIE.5491..551J,2008SPIE.7013E..91R,2008SPIE.7013E..45Z,2008PASP..120..617V,2010A&A...515A..69M,2011A&A...535A..68A,2011PASP..123..964Z}. However, current interferometers lack sufficient accuracy for such a detection. When observing a star with a faint companion, their fringe patterns add up incoherently and the presence of a planet causes a slight change in the phases and, consequently, the closure phases.This difference can be measured with a temporal survey and should be corrected with the intrinsic closure phases of the host star. \\
In this framework, theoretical predictions of the closure phases of the host stars are crucial. Closure phase between three (or more) telescopes is the sum of all phase differences: this procedure removes the atmospheric contribution, leaving the phase information of
the object visibility unaltered. The major biases or systematic errors of closure phases come from non-closed triangles introduced in the measurement process, which in principle, can be precisely calibrated. Therefore, it is a good observable for stable and precise measurements
 \citep[e.g.][]{2007NewAR..51..604M}. The closure phase thus offers an important complementary piece of information, revealing asymmetries and inhomogeneities of stellar disk images.

\subsection{Closure phases from the star alone}

  \begin{figure}
   \centering
   \begin{tabular}{ccc}
   \includegraphics[width=0.98\hsize]{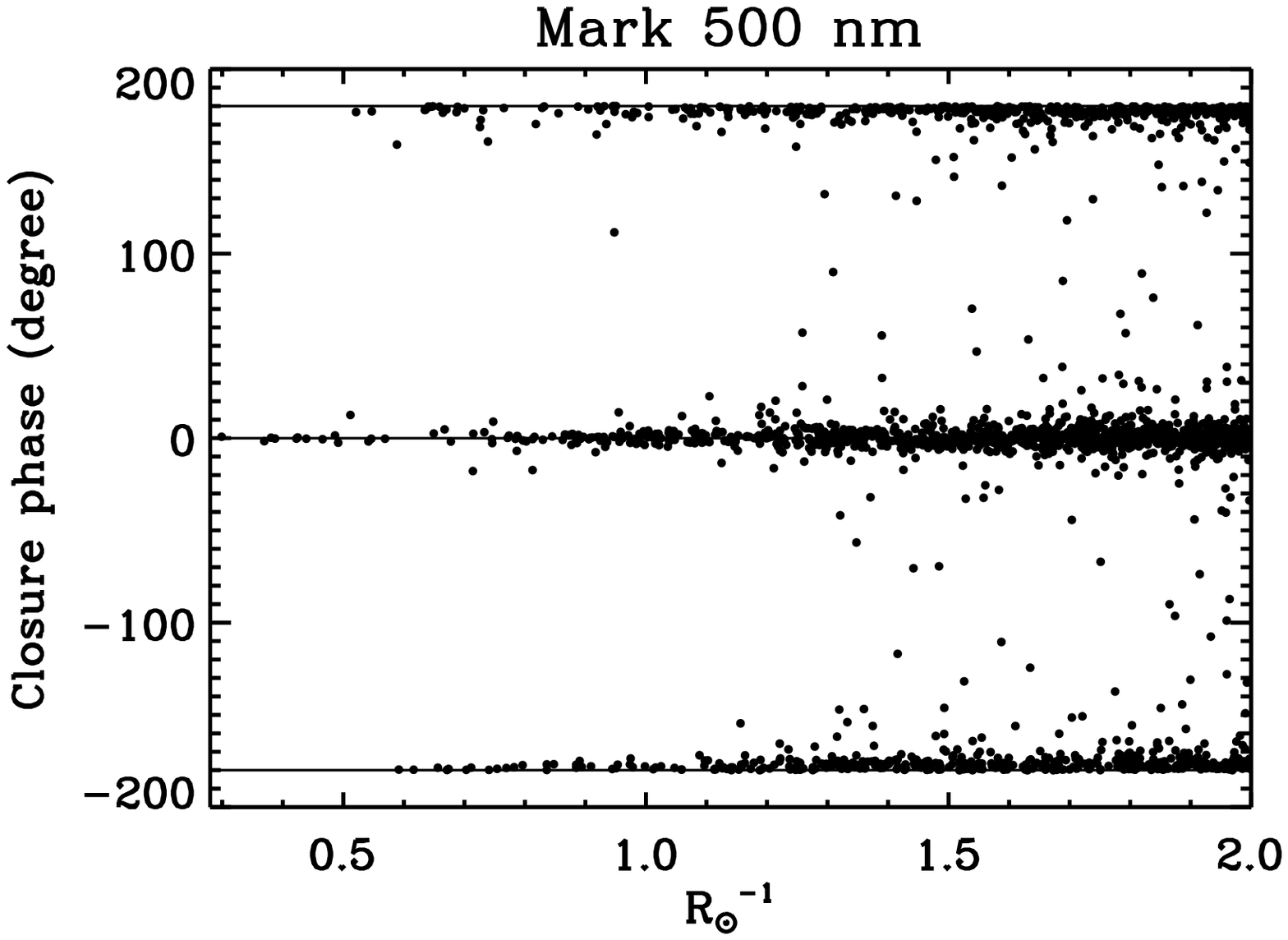}\\
       \includegraphics[width=0.98\hsize]{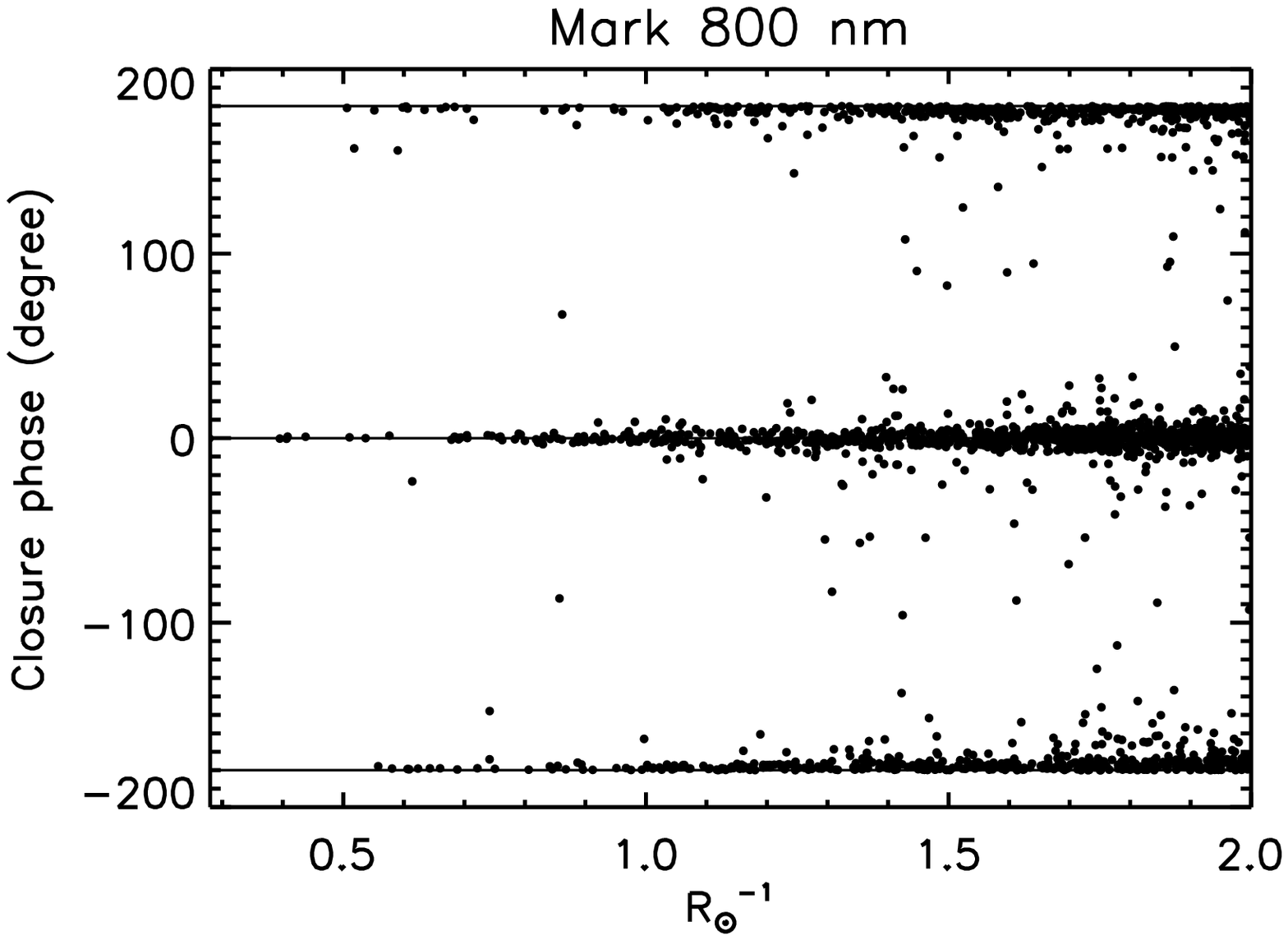}\\
           \includegraphics[width=0.98\hsize]{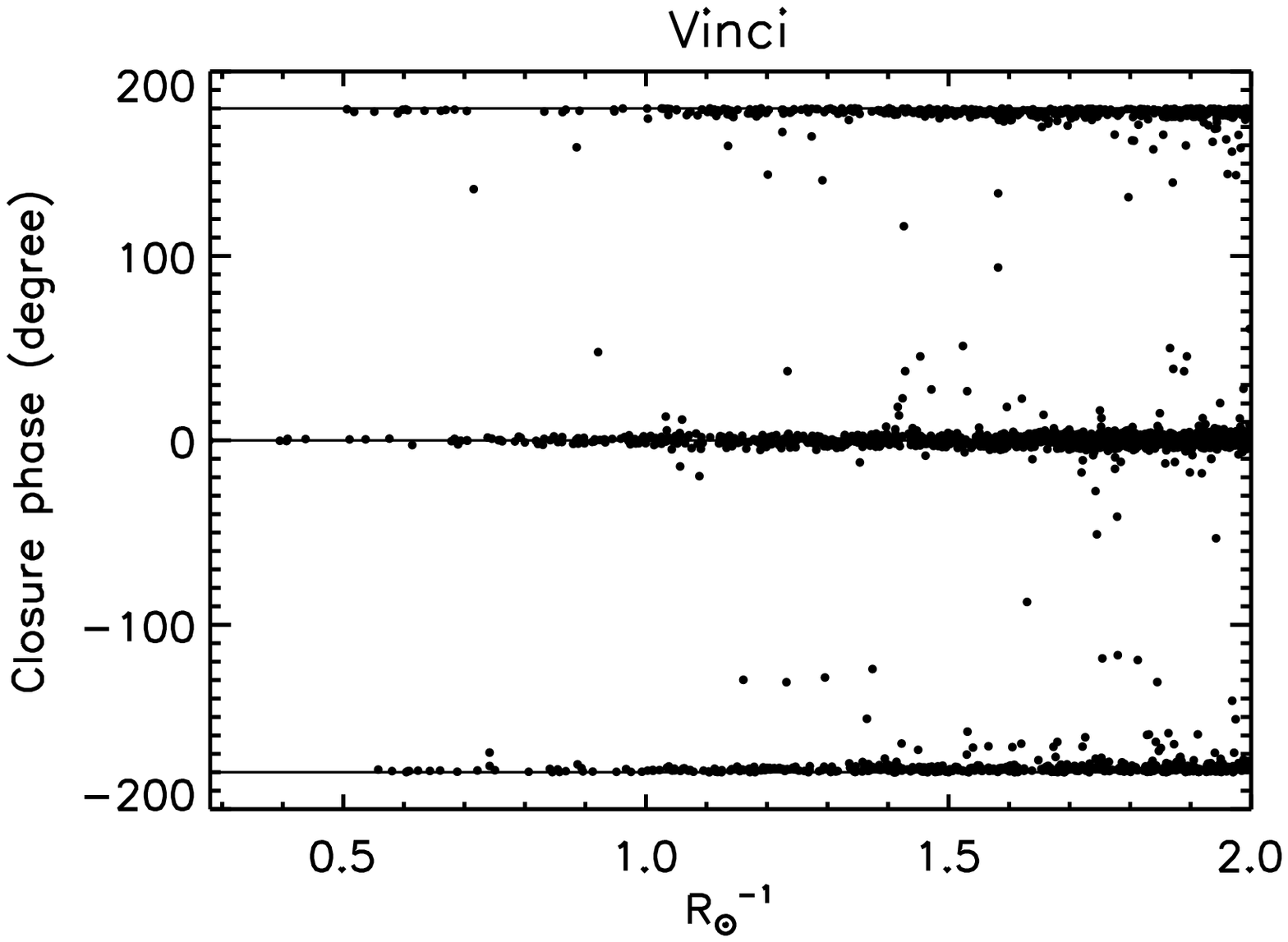}
        \end{tabular}
      \caption{Scatter plot of closure phases of 2000 random baseline triangles
with a maximum linear extension of $\sim$155 m at 500 nm, $\sim$250 m at 800 nm, and $\sim$680 m at 2.2 $\mu$m, respectively. A parallax of 284.56 mas \citep{vanleeuwen07} and an apparent radius of 5.390 mas have been used.}
        \label{closure}
   \end{figure}

Figure~\ref{closure} displays deviations from the axisymmetric case (zero or $\pm\pi$) that are particularly occurring in optical filters. There is a correlation between Fig.~\ref{closure} and Fig.~\ref{visibility} because the scatter of closure phases increases with spatial frequencies as for the visibilities: smaller structures need large baselines to be resolved. Moreover, it is visible that the closure phase signal becomes important for frequencies larger than $\sim0.6$~R$^{-1}_\odot$ (top of the third visibility lobe, i.e., $vis\sim0.06$ from Fig.~\ref{visibility}). These predictions can be constrained by the level of asymmetry and inhomogeneity of stellar disks by accumulating observations on closure phase at short and long baselines. \\
Observing dwarf stars at high spatial resolution is thus crucial to characterize the granulation pattern using closure phases. This requires
 observations at high spatial frequencies (from 3rd lobe on) and especially in the optical range. In fact, considering that the maximum
 baseline of CHARA is 331 meters \citep{2005ApJ...628..453T}, for a star like Procyon with a parallax of 284.56 mas \citep{vanleeuwen07}, a baseline of
 $\sim$55 (240) meters is necessary to probe the third lobe at 0.5 (2.2) $\mu$m. For comparison, the nominal measurement error with CHARA array is 0.3$^\circ$ with a peak of performance of 0.1$^\circ$ for a shorter triangle \cite{2008SPIE.7013E..45Z} (to be compared with closure phases value of Fig.~\ref{closure}).
 
 \subsection{Closure phases from the hosting star plus the hot Jupiter companion}
 
 Among the detected exoplanets, the direct detection and the characterization of their atmospheres appears currently within reach for very close planet-star system ($<$0.1AU) and for planets with temperatures $\gtrsim$1000 K, implying their infrared flux is $\sim10^{-3}$ of their host stars. Since the bulk of the energy from hot Jupiters emerges from the near-infrared between 1-3~$\mu$m \citep{2008ApJ...678.1436B}, interferometry in the near infrared band (like in the range of the {\sc Vinci} filter centered at $\sim2.2$ $\mu$m) can provide measurements capable to detect and characterize the planets. Detecting hot Jupiters from Earth is endeavoring because it is challenging and at the limits of current performance of interferometry and, once these conditions are met, the signal from the host stars must also be known in detail. As visible in the synthetic stellar disk images of Fig.~\ref{images}, the granulation is a non-negligible aspect of the surface of dwarf stars and has an important signal in the closure phases (Fig.~\ref{closure}).
 
 To estimate the impact of the granulation noise on the planet detection, we used the RHD simulation of Procyon and added a virtual companion to the star. 
 The modelling of the flux of an irradiated planet requires careful attention on the radiative transfer conditions related to the stellar irradiation and therefore we used the models of \cite{2001ApJ...556..885B}. In particular, we used spectra of hot irradiated extrasolar planet around a star with about the same spectral type of Procyon, a mass of 1 Jupiter mass, and an intrinsic temperature 1000 K. We assumed a radius of 1.2 Jupiter radii and various orbital distances [0.1, 0.25, 0.5, 1.0] AU following \cite{2010A&A...515A..69M}. The atmospheric composition of the hot Jupiter is identical to the two models of \cite{2001ApJ...556..357A} where: (1) dust (particles and grains) remains in the upper atmosphere and (2) where dust has been removed from the upper atmosphere by condensation and gravitational settling.
 
 \begin{figure}
   \centering
   \includegraphics[width=1.0\hsize]{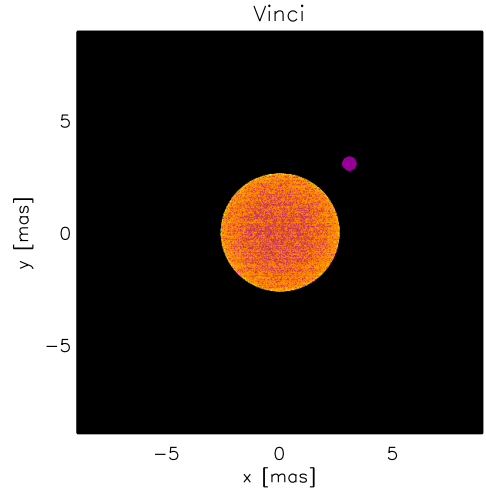}\\
      \caption{Synthetic stellar disk images (yellow-orange scale) of the simulation in {\sc Vinci} filter (the intensity range is [0.3 - 0.9$\times10^5$]\,erg\,cm$^{-2}$\,s$^{-1}$\,{\AA}$^{-1}$) together with a hot Jupiter (pink) at a distance of 0.1 AU for a star with a parallax of 284.56 mas \citep{vanleeuwen07}. The intensity of the hot Jupiter is $\sim$11 times lower than the maximum stellar surface intensity.}
        \label{planet1}
   \end{figure}
 
Figure~\ref{planet1} shows the geometrical configuration of the star-planet system for a particular distance. First we average the exoplanetary spectrum in the range of the {\sc Mark}~III 800 nm and {\sc Vinci} filters (we had no data for the exoplanet spectrum in the range of {\sc Mark}~III 500 nm filter), and then we used this intensity for the stellar companion in the Fig.~\ref{planet1} . The intensity of the planet is stronger in the infrared with respect to the optical. The ratio between the stellar intensity at its center (i.e., $\mu=1$), $I_{\rm{{\sc Vinci}}}$ or $I_{\rm{{\sc Mark}800}}$, and the planet integrated intensity in the same filter, $I_{\rm{planet}}$, is:
\begin{eqnarray}
  \frac{I_{\rm{{\sc Vinci}}}}{I_{\rm{planet}}} &=&11.5, 31.7, 42.7, 46.3, \\
  \frac{I_{\rm{{\sc Mark} 800}}}{I_{\rm{planet}}} &=& 34.7, 104.0, 130.0, 182.5 
  \end{eqnarray}
   for distances [0.1, 0.25, 0.5, 1.0] AU, respectively. 
   
\begin{figure}
   \centering
   \begin{tabular}{cc}
      \includegraphics[width=1.0\hsize]{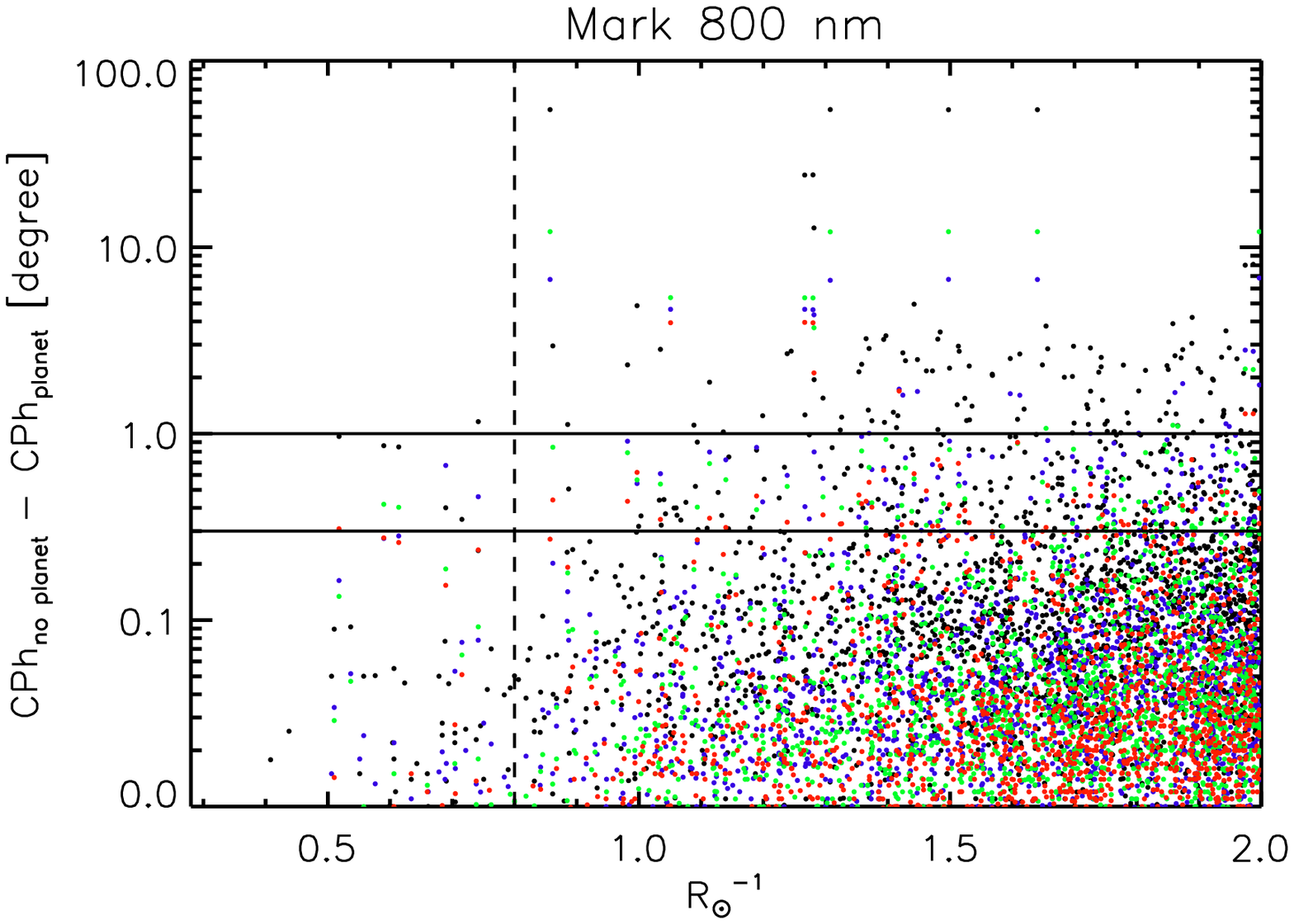}\\
                                       \includegraphics[width=1.0\hsize]{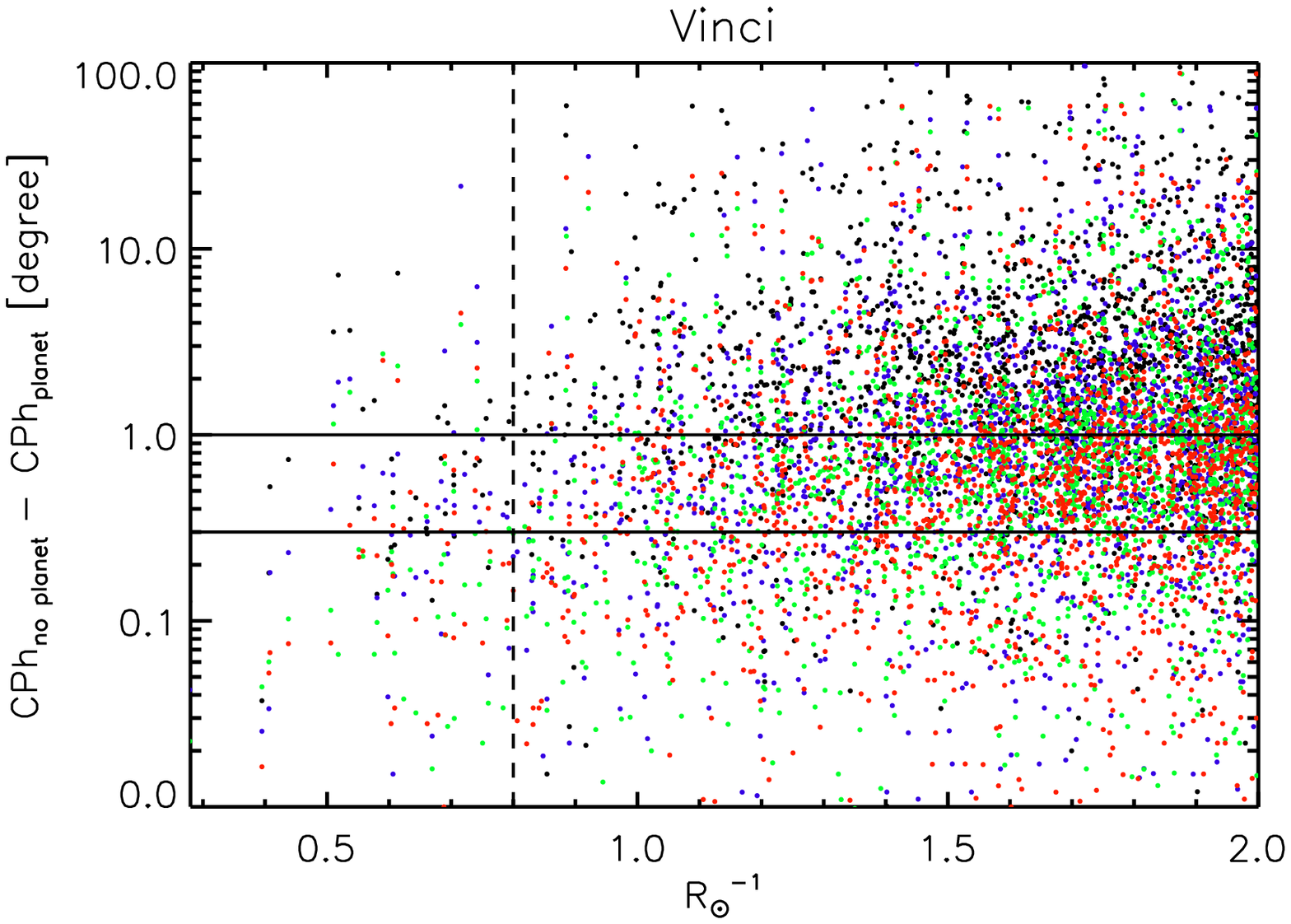}
                     \end{tabular}
      \caption{Absolute differences of the closure phases computed for the 2000 random baseline triangles in the case of a system without the presence of a hot Jupiter companion ($CPh_{\rm{no planet}}$) and a system star-planet ($CPh_{\rm{planet}}$, see text and Fig.~\ref{planet1}). The horizontal solid lines indicate the 1.0 and 0.3 degrees and the vertical dashed line the spatial frequency corresponding to the end of third visibility lobe. The black, blue, green, and red colors denote the planet-star distance of [0.1, 0.25, 0.5, 1.0] AU, respectively. The maximum linear extension, the apparent radius and the parallax are the same as Fig.~\ref{closure}.
      }
        \label{planet2}
   \end{figure}

We computed the closure phases from the image of Fig.~\ref{planet1} and for similar systems corresponding to the {\sc Vinci} and {\sc Mark}~III 800 nm filters and for the distances reported above. These closure phases were compared with the resulting phases computed for exactly the same triangles but for a system without the presence of a planet. Figure~\ref{planet2} displays the absolute differences between closure phases with and without the presence of a hot Jupiter. Setting as a reference the closure phase nominal error of CHARA (0.3$^\circ$) and also 1$^\circ$ (horizontal lines in the plot), only the {\sc Vinci} filter gives differences that should be detectable on the third lobe ($0.5\lesssim R_{\odot}^{-1}\lesssim0.8$) while there is no signature on the second lobe ($R_{\odot}^{-1}\lesssim0.5$). The absolute difference increases as a function of spatial frequency. It is indistinguishable for all the planet' distances for frequencies larger than 0.8$\sim R_{\odot}^{-1}$.
 
The purpose of this Section was to show the alteration of the signal due to the granulation noise in the detection of a hot Jupiter around Procyon-like stars using closure phases. So far, no companions have been detected around Procyon except for a white dwarf astrometric companion detected already in the 19th century (see Section~\ref{sectintro}).


The studies of hot Jupiter atmospheres will reveal their composition, structure, dynamics, and planet formation processes. It is then very important to have a complete knowledge of the host star to reach these aims.

\section{Conclusions}

We have provided new predictions of interferometric and spectroscopic observables for Procyon, based on RHD hydrodynamical simulation, that affect the fundamental parameter determination of the star and are important for the detection hot Jupiter exoplanets.\\

We have studied the impact of the granulation pattern on the center-to-limb intensity profiles and provided limb-darkening coefficients in the optical as well as in the infrared. We showed that synthetic visibility curves from RHD simulation are systematically lower than uniform disk and this effect is stronger in the optical filters. In addition to this, visibilities display fluctuations increasing with spatial frequencies (i.e., departure from circular symmetry) that becomes $\sim2\%$ on the top of third lobe in {\sc Mark}~III 500 nm filter. However, in the {\sc Mark}~III 800 nm and {\sc Vinci} filter the dispersion is much weaker. \\

We have derived new angular diameters at different wavelengths with two independent methods based on the RHD simulation. The angular diameter in the {\sc Vinci} filter is $\theta=5.390$ mas, which leads to an effective temperature of $\teff = 6591 \pm 43 $ K or $6556\pm 84$ K, depending on the bolometric flux considered. This value is now more consistent with $\teffir= 6621 \pm 80$ K from the infrared flux method \citep[][and Casagrande private communication]{2010A&A...512A..54C}. \\
Using an independent estimation of the radius from asteroseismology, we found $\theta_{\rm seismic} = 5.36 \pm 0.07$ mas. This radius agrees well with our interferometric value within the error bars. \\
Eventually, the combination of the astrometric mass and our new interferometric diameter leads to a new gravity $\logg = 4.01 \pm 0.03$ [cm/s$^2$], which is  larger by $0.05$ dex than the value derived in \citet{2002ApJ...567..544A}. \\

We provided accurate comparison of synthetic spectrum from a RHD simulation to observations from the ultraviolet to the infrared. The photometric $BVRI$ colors are very similar to the observations while it is difficult to conclude on the infrared colors because of the saturation of the observations. Also the comparison with the absolute spectrophotometric measurements, collected in \citet{2005ApJ...633..424A}, is in agreement. We conclude that the mean thermal gradient of the simulation, reflected by the spectral energy distribution, is in very good agreement with Procyon.\\

The convective related surface structures impact also the signal of the closure phases that show departures from symmetry at about the same spatial frequencies of visibility curves. We concluded that closure phases not equal to 0 or $\pm\pi$ may be detected with today interferometers such as CHARA in the visible filters where the baselines are long enough to get to the second/third lobes. \\
We estimated the impact of the granulation noise on the hot Jupiter detection using closure phases around stars with the same spectral type of Procyon. We used the synthetic stellar disks obtained from RHD in the infrared and optical filters and added a virtual companion to the star based on real integrated spectra of irradiated extrasolar planet. Then, we computed the closure phases for planet-star system and star only and found that there is a non-negligible and detectable contamination to the signal of the hot Jupiter due to the granulation from spatial frequencies longward of the third lobe. This is valid only for the infrared where the energy from the hot Jupiters is stronger. It is thus very important to have a comprehensive knowledge of the host star to detect and characterize hot Jupiters, and RHD simulations are very important to reach this aim. In a forthcoming paper, we will extend this analysis to solar type stars and K giants across the HR diagram.

\begin{acknowledgements}
The authors thank a lot J. Aufdenberg for his help and the enlightening discussions. A.C. is supported in part by an {\it Action de recherche concert\'ee} (ARC) grant from the {\it Direction g\'en\'erale de l'Enseignement non obligatoire et de la Recherche scientifique - Direction de la Recherche scientifique - Communaut\'e fran\c{c}aise de Belgique}. A.C. is also supported by the F.R.S.-FNRS FRFC grant 2.4513.11. We thank the Rechenzentrum Garching (RZG) for providing the computational resources necessary for this
work. This research received the support of PHASE, the high angular resolution partnership between ONERA, Observatoire de Paris, CNRS and University Denis Diderot Paris 7.
\end{acknowledgements}

\bibliographystyle{aa}
\bibliography{biblio}

\begin{thebibliography}{90}
\expandafter\ifx\csname natexlab\endcsname\relax\def\natexlab#1{#1}\fi
\expandafter\ifx\csname url\endcsname\relax
  \def\url#1{{\tt #1}}\fi
\expandafter\ifx\csname urlprefix\endcsname\relax\def\urlprefix{URL }\fi

\bibitem[{{Absil} et~al.(2011){Absil}, {Le Bouquin}, {Berger}
  et~al.}]{2011A&A...535A..68A}
{Absil} O., {Le Bouquin} J.B., {Berger} J.P., et~al., Nov. 2011, \aap, 535, A68

\bibitem[{{Allard} et~al.(2001){Allard}, {Hauschildt}, {Alexander}, {Tamanai},
  \& {Schweitzer}}]{2001ApJ...556..357A}
{Allard} F., {Hauschildt} P.H., {Alexander} D.R., {Tamanai} A., {Schweitzer}
  A., Jul. 2001, \apj, 556, 357

\bibitem[{{Allende Prieto} et~al.(2002){Allende Prieto}, {Asplund},
  {Garc{\'{\i}}a L{\'o}pez}, \& {Lambert}}]{2002ApJ...567..544A}
{Allende Prieto} C., {Asplund} M., {Garc{\'{\i}}a L{\'o}pez} R.J., {Lambert}
  D.L., Mar. 2002, \apj, 567, 544

\bibitem[{{Arentoft} et~al.(2008){Arentoft}, {Kjeldsen}, {Bedding}
  et~al.}]{arentoft08}
{Arentoft} T., {Kjeldsen} H., {Bedding} T.R., et~al., Nov. 2008, \apj, 687,
  1180

\bibitem[{{Asplund} et~al.(2009){Asplund}, {Grevesse}, {Sauval}, \&
  {Scott}}]{asplund09}
{Asplund} M., {Grevesse} N., {Sauval} A.J., {Scott} P., Sep. 2009, \araa, 47,
  481

\bibitem[{{Atroshchenko} et~al.(1989){Atroshchenko}, {Gadun}, \&
  {Kostik}}]{atroshchenko89}
{Atroshchenko} I.N., {Gadun} A.S., {Kostik} R.I., 1989, In: {R.~J.~Rutten \&
  G.~Severino} (ed.) NATO ASIC Proc. 263: Solar and Stellar Granulation, 521

\bibitem[{{Aufdenberg} et~al.(2005){Aufdenberg}, {Ludwig}, \&
  {Kervella}}]{2005ApJ...633..424A}
{Aufdenberg} J.P., {Ludwig} H.G., {Kervella} P., Nov. 2005, \apj, 633, 424

\bibitem[{{Auwers}(1862)}]{auwers1862}
{Auwers} G.F.J.A., 1862, {De motu proprio Procyonis variabili ...}

\bibitem[{{Barban} et~al.(1999){Barban}, {Michel}, {Martic} et~al.}]{barban99}
{Barban} C., {Michel} E., {Martic} M., et~al., Oct. 1999, \aap, 350, 617

\bibitem[{{Barman} et~al.(2001){Barman}, {Hauschildt}, \&
  {Allard}}]{2001ApJ...556..885B}
{Barman} T.S., {Hauschildt} P.H., {Allard} F., Aug. 2001, \apj, 556, 885

\bibitem[{{Bedding} \& {Kjeldsen}(2003)}]{bedding03}
{Bedding} T.R., {Kjeldsen} H., 2003, \pasa, 20, 203

\bibitem[{{Bedding} et~al.(2010){Bedding}, {Kjeldsen}, {Campante}
  et~al.}]{bedding10}
{Bedding} T.R., {Kjeldsen} H., {Campante} T.L., et~al., Apr. 2010, \apj, 713,
  935

\bibitem[{{Belkacem} et~al.(2011){Belkacem}, {Goupil}, {Dupret}
  et~al.}]{belkacem11}
{Belkacem} K., {Goupil} M.J., {Dupret} M.A., et~al., Jun. 2011, \aap, 530, A142

\bibitem[{{Bessel}(1844)}]{bessel1844}
{Bessel} F.W., Dec. 1844, \mnras, 6, 136

\bibitem[{{Bessel}(1990)}]{1990A&AS...83..357B}
{Bessel} M.S., May 1990, \aaps, 83, 357

\bibitem[{{Bessell}(1990)}]{1990PASP..102.1181B}
{Bessell} M.S., Oct. 1990, \pasp, 102, 1181

\bibitem[{{Bigot} et~al.(2006){Bigot}, {Kervella}, {Th{\'e}venin}, \&
  {S{\'e}gransan}}]{2006A&A...446..635B}
{Bigot} L., {Kervella} P., {Th{\'e}venin} F., {S{\'e}gransan} D., Feb. 2006,
  \aap, 446, 635

\bibitem[{{Bigot} et~al.(2011){Bigot}, {Mourard}, {Berio}
  et~al.}]{2011A&A...534L...3B}
{Bigot} L., {Mourard} D., {Berio} P., et~al., Oct. 2011, \aap, 534, L3

\bibitem[{{Bonanno} et~al.(2007){Bonanno}, {K{\"u}ker}, \&
  {Patern{\`o}}}]{bonanno07}
{Bonanno} A., {K{\"u}ker} M., {Patern{\`o}} L., Feb. 2007, \aap, 462, 1031

\bibitem[{{Brown} et~al.(1991){Brown}, {Gilliland}, {Noyes}, \&
  {Ramsey}}]{brown91}
{Brown} T.M., {Gilliland} R.L., {Noyes} R.W., {Ramsey} L.W., Feb. 1991, \apj,
  368, 599

\bibitem[{{Burrows} et~al.(2008){Burrows}, {Budaj}, \&
  {Hubeny}}]{2008ApJ...678.1436B}
{Burrows} A., {Budaj} J., {Hubeny} I., May 2008, \apj, 678, 1436

\bibitem[{{Casagrande} et~al.(2010){Casagrande}, {Ram{\'{\i}}rez},
  {Mel{\'e}ndez}, {Bessell}, \& {Asplund}}]{2010A&A...512A..54C}
{Casagrande} L., {Ram{\'{\i}}rez} I., {Mel{\'e}ndez} J., {Bessell} M.,
  {Asplund} M., Mar. 2010, \aap, 512, A54

\bibitem[{{Chiavassa} et~al.(2009){Chiavassa}, {Plez}, {Josselin}, \&
  {Freytag}}]{2009A&A...506.1351C}
{Chiavassa} A., {Plez} B., {Josselin} E., {Freytag} B., Nov. 2009, \aap, 506,
  1351

\bibitem[{{Chiavassa} et~al.(2010{\natexlab{a}}){Chiavassa}, {Collet},
  {Casagrande}, \& {Asplund}}]{2010A&A...524A..93C}
{Chiavassa} A., {Collet} R., {Casagrande} L., {Asplund} M., Dec.
  2010{\natexlab{a}}, \aap, 524, A93

\bibitem[{{Chiavassa} et~al.(2010{\natexlab{b}}){Chiavassa}, {Haubois}, {Young}
  et~al.}]{2010A&A...515A..12C}
{Chiavassa} A., {Haubois} X., {Young} J.S., et~al., Jun. 2010{\natexlab{b}},
  \aap, 515, A12

\bibitem[{{Chollet} \& {Sinceac}(1999)}]{chollet99}
{Chollet} F., {Sinceac} V., Oct. 1999, \aaps, 139, 219

\bibitem[{{Code} et~al.(1976){Code}, {Bless}, {Davis}, \& {Brown}}]{code76}
{Code} A.D., {Bless} R.C., {Davis} J., {Brown} R.H., Jan. 1976, \apj, 203, 417

\bibitem[{{Collet} et~al.(2011){Collet}, {Magic}, \&
  {Asplund}}]{2011JPhCS.328a2003C}
{Collet} R., {Magic} Z., {Asplund} M., Dec. 2011, Journal of Physics Conference
  Series, 328, 012003

\bibitem[{{di Mauro} \& {Christensen-Dalsgaard}(2001)}]{dimauro01}
{di Mauro} M.P., {Christensen-Dalsgaard} J., 2001, In: {P.~Brekke, B.~Fleck, \&
  J.~B.~Gurman} (ed.) Recent Insights into the Physics of the Sun and
  Heliosphere: Highlights from SOHO and Other Space Missions, vol. 203 of IAU
  Symposium, 94

\bibitem[{{Dravins}(1987)}]{dravins87}
{Dravins} D., Jan. 1987, \aap, 172, 211

\bibitem[{{Eggen} \& {Greenstein}(1965)}]{eggen65}
{Eggen} O.J., {Greenstein} J.L., Jan. 1965, \apj, 141, 83

\bibitem[{{Eggenberger} et~al.(2004){Eggenberger}, {Carrier}, {Bouchy}, \&
  {Blecha}}]{eggenberger04}
{Eggenberger} P., {Carrier} F., {Bouchy} F., {Blecha} A., Jul. 2004, \aap, 422,
  247

\bibitem[{{Eggenberger} et~al.(2005){Eggenberger}, {Carrier}, \&
  {Bouchy}}]{eggenberger05}
{Eggenberger} P., {Carrier} F., {Bouchy} F., Jan. 2005, \na, 10, 195

\bibitem[{{Freytag} et~al.(2002){Freytag}, {Steffen}, \&
  {Dorch}}]{2002AN....323..213F}
{Freytag} B., {Steffen} M., {Dorch} B., Jul. 2002, Astronomische Nachrichten,
  323, 213

\bibitem[{{Freytag} et~al.(2011){Freytag}, {Steffen}, {Ludwig}
  et~al.}]{2011arXiv1110.6844F}
{Freytag} B., {Steffen} M., {Ludwig} H.G., et~al., Oct. 2011, ArXiv e-prints

\bibitem[{{Fuhrmann} et~al.(1997){Fuhrmann}, {Pfeiffer}, {Frank}, {Reetz}, \&
  {Gehren}}]{fuhrmann97}
{Fuhrmann} K., {Pfeiffer} M., {Frank} C., {Reetz} J., {Gehren} T., Jul. 1997,
  \aap, 323, 909

\bibitem[{{Gatewood} \& {Han}(2006)}]{gatewood06}
{Gatewood} G., {Han} I., Feb. 2006, \aj, 131, 1015

\bibitem[{{Gelly} et~al.(1986){Gelly}, {Grec}, \& {Fossat}}]{gelly86}
{Gelly} B., {Grec} G., {Fossat} E., Aug. 1986, \aap, 164, 383

\bibitem[{{Gelly} et~al.(1988){Gelly}, {Grec}, \& {Fossat}}]{gelly88}
{Gelly} B., {Grec} G., {Fossat} E., 1988, In: {J.~Christensen-Dalsgaard \&
  S.~Frandsen} (ed.) Advances in Helio- and Asteroseismology, vol. 123 of IAU
  Symposium, 249

\bibitem[{{Girard} et~al.(2000){Girard}, {Wu}, {Lee} et~al.}]{girard00}
{Girard} T.M., {Wu} H., {Lee} J.T., et~al., May 2000, \aj, 119, 2428

\bibitem[{{Glushneva} et~al.(1992){Glushneva}, {Kharitonov}, {Kniazeva}, \&
  {Shenavrin}}]{1992A&AS...92....1G}
{Glushneva} I.N., {Kharitonov} A.V., {Kniazeva} L.N., {Shenavrin} V.I., Jan.
  1992, \aaps, 92, 1

\bibitem[{{Gray}(1967)}]{gray67}
{Gray} D.F., Aug. 1967, \apj, 149, 317

\bibitem[{{Gray}(1981)}]{gray81}
{Gray} D.F., Dec. 1981, \apj, 251, 583

\bibitem[{{Griffin}(1971)}]{griffin71}
{Griffin} R., 1971, \mnras, 155, 139

\bibitem[{{Guenther} \& {Demarque}(1993)}]{guenther93}
{Guenther} D.B., {Demarque} P., Mar. 1993, \apj, 405, 298

\bibitem[{{Guenther} et~al.(2008){Guenther}, {Kallinger}, {Gruberbauer}
  et~al.}]{guenther08}
{Guenther} D.B., {Kallinger} T., {Gruberbauer} M., et~al., Nov. 2008, \apj,
  687, 1448

\bibitem[{{Gustafsson} et~al.(2008){Gustafsson}, {Edvardsson}, {Eriksson}
  et~al.}]{2008A&A...486..951G}
{Gustafsson} B., {Edvardsson} B., {Eriksson} K., et~al., Aug. 2008, \aap, 486,
  951

\bibitem[{{Hanbury Brown} et~al.(1967){Hanbury Brown}, {Davis}, {Allen}, \&
  {Rome}}]{hanbury67}
{Hanbury Brown} R., {Davis} J., {Allen} L.R., {Rome} J.M., 1967, \mnras, 137,
  393

\bibitem[{{Hanbury Brown} et~al.(1974){Hanbury Brown}, {Davis}, {Lake}, \&
  {Thompson}}]{1974MNRAS.167..475H}
{Hanbury Brown} R., {Davis} J., {Lake} R.J.W., {Thompson} R.J., Jun. 1974,
  \mnras, 167, 475

\bibitem[{{Hartmann} et~al.(1975){Hartmann}, {Garrison}, \&
  {Katz}}]{hartmann75}
{Hartmann} L., {Garrison} L.M. Jr., {Katz} A., Jul. 1975, \apj, 199, 127

\bibitem[{{Hayek} et~al.(2010){Hayek}, {Asplund}, {Carlsson}
  et~al.}]{2010A&A...517A..49H}
{Hayek} W., {Asplund} M., {Carlsson} M., et~al., Jul. 2010, \aap, 517, A49

\bibitem[{{Joergens} \& {Quirrenbach}(2004)}]{2004SPIE.5491..551J}
{Joergens} V., {Quirrenbach} A., Oct. 2004, In: {W.~A.~Traub} (ed.) SPIE Proc.,
  vol. 5491, 551

\bibitem[{{Kervella} et~al.(2003{\natexlab{a}}){Kervella}, {Gitton},
  {Segransan} et~al.}]{2003SPIE.4838..858K}
{Kervella} P., {Gitton} P.B., {Segransan} D., et~al., Feb. 2003{\natexlab{a}},
  In: {W.~A.~Traub} (ed.) SPIE Proc., vol. 4838, 858--869

\bibitem[{{Kervella} et~al.(2003{\natexlab{b}}){Kervella}, {Th{\'e}venin},
  {Morel}, {Bord{\'e}}, \& {Di Folco}}]{2003A&A...408..681K}
{Kervella} P., {Th{\'e}venin} F., {Morel} P., {Bord{\'e}} P., {Di Folco} E.,
  Sep. 2003{\natexlab{b}}, \aap, 408, 681

\bibitem[{{Kervella} et~al.(2003{\natexlab{c}}){Kervella}, {Th{\'e}venin},
  {S{\'e}gransan} et~al.}]{2003A&A...404.1087K}
{Kervella} P., {Th{\'e}venin} F., {S{\'e}gransan} D., et~al., Jun.
  2003{\natexlab{c}}, \aap, 404, 1087

\bibitem[{{Kervella} et~al.(2004{\natexlab{a}}){Kervella}, {S{\'e}gransan}, \&
  {Coud{\'e} du Foresto}}]{2004A&A...425.1161K}
{Kervella} P., {S{\'e}gransan} D., {Coud{\'e} du Foresto} V., Oct.
  2004{\natexlab{a}}, \aap, 425, 1161

\bibitem[{{Kervella} et~al.(2004{\natexlab{b}}){Kervella}, {Th{\'e}venin},
  {Morel} et~al.}]{2004A&A...413..251K}
{Kervella} P., {Th{\'e}venin} F., {Morel} P., et~al., Jan. 2004{\natexlab{b}},
  \aap, 413, 251

\bibitem[{{Kjeldsen} \& {Bedding}(2011)}]{kjeldsen11}
{Kjeldsen} H., {Bedding} T.R., May 2011, \aap, 529, L8

\bibitem[{{Lopez} et~al.(2000){Lopez}, {Petrov}, \&
  {Vannier}}]{2000SPIE.4006..407L}
{Lopez} B., {Petrov} R.G., {Vannier} M., Jul. 2000, In: {P.~L{\'e}na \&
  A.~Quirrenbach} (ed.) SPIE Proc., vol. 4006, 407--411

\bibitem[{{Marti{\'c}} et~al.(1999){Marti{\'c}}, {Schmitt}, {Lebrun}
  et~al.}]{Martic99}
{Marti{\'c}} M., {Schmitt} J., {Lebrun} J.C., et~al., Nov. 1999, \aap, 351, 993

\bibitem[{{Marti{\'c}} et~al.(2004){Marti{\'c}}, {Lebrun}, {Appourchaux}, \&
  {Korzennik}}]{martic04}
{Marti{\'c}} M., {Lebrun} J.C., {Appourchaux} T., {Korzennik} S.G., Apr. 2004,
  \aap, 418, 295

\bibitem[{{Matter} et~al.(2010){Matter}, {Vannier}, {Morel}
  et~al.}]{2010A&A...515A..69M}
{Matter} A., {Vannier} M., {Morel} S., et~al., Jun. 2010, \aap, 515, A69

\bibitem[{{Matthews} et~al.(2004){Matthews}, {Kuschnig}, {Guenther}
  et~al.}]{matthews04}
{Matthews} J.M., {Kuschnig} R., {Guenther} D.B., et~al., Jul. 2004, \nat, 430,
  51

\bibitem[{{Mihalas} et~al.(1988){Mihalas}, {Dappen}, \& {Hummer}}]{MHD}
{Mihalas} D., {Dappen} W., {Hummer} D.G., Aug. 1988, \apj, 331, 815

\bibitem[{{Monnier}(2007)}]{2007NewAR..51..604M}
{Monnier} J.D., Oct. 2007, New Astronomy Review, 51, 604

\bibitem[{{Mosser} et~al.(2008){Mosser}, {Bouchy}, {Marti{\'c}}
  et~al.}]{mosser08}
{Mosser} B., {Bouchy} F., {Marti{\'c}} M., et~al., Jan. 2008, \aap, 478, 197

\bibitem[{{Mourard} et~al.(2009){Mourard}, {Clausse}, {Marcotto}
  et~al.}]{2009A&A...508.1073M}
{Mourard} D., {Clausse} J.M., {Marcotto} A., et~al., Dec. 2009, \aap, 508, 1073

\bibitem[{{Mozurkewich} et~al.(1991){Mozurkewich}, {Johnston}, {Simon}
  et~al.}]{1991AJ....101.2207M}
{Mozurkewich} D., {Johnston} K.J., {Simon} R.S., et~al., Jun. 1991, \aj, 101,
  2207

\bibitem[{{Nelson}(1980)}]{nelson80}
{Nelson} G.D., Jun. 1980, \apj, 238, 659

\bibitem[{{Nordlund}(1982)}]{1982A&A...107....1N}
{Nordlund} A., Mar. 1982, \aap, 107, 1

\bibitem[{{Nordlund} \& {Dravins}(1990)}]{nordlund90}
{Nordlund} A., {Dravins} D., Feb. 1990, \aap, 228, 155

\bibitem[{{Nordlund} et~al.(2009){Nordlund}, {Stein}, \&
  {Asplund}}]{2009LRSP....6....2N}
{Nordlund} {\AA}., {Stein} R.F., {Asplund} M., Apr. 2009, Living Reviews in
  Solar Physics, 6, 2

\bibitem[{{Provencal} et~al.(2002){Provencal}, {Shipman}, {Koester},
  {Wesemael}, \& {Bergeron}}]{provencal02}
{Provencal} J.L., {Shipman} H.L., {Koester} D., {Wesemael} F., {Bergeron} P.,
  Mar. 2002, \apj, 568, 324

\bibitem[{{Provost} et~al.(2006){Provost}, {Berthomieu}, {Marti{\'c}}, \&
  {Morel}}]{provost06}
{Provost} J., {Berthomieu} G., {Marti{\'c}} M., {Morel} P., Dec. 2006, \aap,
  460, 759

\bibitem[{{Quirrenbach}(2001)}]{2001ARA&A..39..353Q}
{Quirrenbach} A., 2001, \araa, 39, 353

\bibitem[{{Renard} et~al.(2008){Renard}, {Absil}, {Berger}
  et~al.}]{2008SPIE.7013E..91R}
{Renard} S., {Absil} O., {Berger} J., et~al., Jul. 2008, In: SPIE Proc., vol.
  7013

\bibitem[{{Rodr{\'{\i}}guez-Pascual} et~al.(1999){Rodr{\'{\i}}guez-Pascual},
  {Gonz{\'a}lez-Riestra}, {Schartel}, \& {Wamsteker}}]{1999A&AS..139..183R}
{Rodr{\'{\i}}guez-Pascual} P.M., {Gonz{\'a}lez-Riestra} R., {Schartel} N.,
  {Wamsteker} W., Oct. 1999, \aaps, 139, 183

\bibitem[{{Schaeberle}(1896)}]{schaeberle1896}
{Schaeberle} J.M., Dec. 1896, \aj, 17, 37

\bibitem[{{Segransan} et~al.(2000){Segransan}, {Beuzit}, {Delfosse}
  et~al.}]{2000SPIE.4006..269S}
{Segransan} D., {Beuzit} J.L., {Delfosse} X., et~al., Jul. 2000, In:
  {P.~L{\'e}na \& A.~Quirrenbach} (ed.) SPIE Proc., vol. 4006, 269--276

\bibitem[{{Shao} et~al.(1988){Shao}, {Colavita}, {Hines}
  et~al.}]{1988A&A...193..357S}
{Shao} M., {Colavita} M.M., {Hines} B.E., et~al., Mar. 1988, \aap, 193, 357

\bibitem[{{Short} \& {Hauschildt}(2005)}]{2005ApJ...618..926S}
{Short} C.I., {Hauschildt} P.H., Jan. 2005, \apj, 618, 926

\bibitem[{{Steffen}(1985)}]{steffen85}
{Steffen} M., Mar. 1985, \aaps, 59, 403

\bibitem[{{Strand}(1951)}]{strand51}
{Strand} K.A., Jan. 1951, \apj, 113, 1

\bibitem[{{ten Brummelaar} et~al.(2005){ten Brummelaar}, {McAlister}, {Ridgway}
  et~al.}]{2005ApJ...628..453T}
{ten Brummelaar} T.A., {McAlister} H.A., {Ridgway} S.T., et~al., Jul. 2005,
  \apj, 628, 453

\bibitem[{{van Belle}(2008)}]{2008PASP..120..617V}
{van Belle} G.T., May 2008, \pasp, 120, 617

\bibitem[{{van Leeuwen}(2007)}]{vanleeuwen07}
{van Leeuwen} F., Nov. 2007, \aap, 474, 653

\bibitem[{{Verner} et~al.(2011){Verner}, {Elsworth}, {Chaplin}
  et~al.}]{verner11}
{Verner} G.A., {Elsworth} Y., {Chaplin} W.J., et~al., Aug. 2011, \mnras, 415,
  3539

\bibitem[{{Wittkowski} et~al.(2004){Wittkowski}, {Aufdenberg}, \&
  {Kervella}}]{2004A&A...413..711W}
{Wittkowski} M., {Aufdenberg} J.P., {Kervella} P., Jan. 2004, \aap, 413, 711

\bibitem[{{Zhao} et~al.(2008){Zhao}, {Monnier}, {ten Brummelaar}, {Pedretti},
  \& {Thureau}}]{2008SPIE.7013E..45Z}
{Zhao} M., {Monnier} J.D., {ten Brummelaar} T., {Pedretti} E., {Thureau} N.D.,
  Jul. 2008, In: Society of Photo-Optical Instrumentation Engineers (SPIE)
  Conference Series, vol. 7013 of Society of Photo-Optical Instrumentation
  Engineers (SPIE) Conference Series

\bibitem[{{Zhao} et~al.(2011){Zhao}, {Monnier}, {Che}
  et~al.}]{2011PASP..123..964Z}
{Zhao} M., {Monnier} J.D., {Che} X., et~al., Aug. 2011, \pasp, 123, 964

\end{thebibliography}

\end{document}